\documentclass[11pt]{article}
\usepackage{adjustbox}
\usepackage{bm}
\usepackage{amsfonts}
\usepackage[colorlinks,citecolor=blue,urlcolor=blue,hyperfootnotes=false]{hyperref}
\usepackage{amssymb}
\usepackage[bottom]{footmisc}
\usepackage[hyperref]{ntheorem}
\usepackage{natbib}
\usepackage{graphicx}
\usepackage{float}
\usepackage[doublespacing,nodisplayskipstretch]{setspace}
\usepackage[margin=1in, a4paper]{geometry}
\usepackage{booktabs}
\usepackage{threeparttable}
\usepackage{amsmath}
\usepackage{lscape}
\usepackage{scrextend}
\usepackage{relsize}
\usepackage{multicol}
\usepackage{chngcntr}
\usepackage{etoolbox}
\usepackage{amssymb}
\usepackage{tabularx}
\usepackage[english]{babel}
\usepackage{multirow}
\usepackage{booktabs}
\usepackage[labelfont=bf]{caption}
\usepackage{float}
\usepackage{titlesec}
\usepackage{array}
\usepackage{color}
\usepackage{afterpage}
\usepackage{framed}
\usepackage{float}
\usepackage[titletoc,title]{appendix}
\usepackage{afterpage}

\allowdisplaybreaks

\providecommand{\U}[1]{\protect\rule{.1in}{.1in}}
\theoremstyle{plain}
\newtheorem{theorem}{Theorem}
\newtheorem{corollary}{Corollary}

\newtheorem{lemma}{Lemma}

\newtheorem{assumption}{Assumption}

\theorembodyfont{\normalfont}
\newtheorem{Rem}{Remark}
\deffootnote{1em}{1.6em}{\thefootnotemark \enskip}

\newcommand\independent{\protect\mathpalette{\protect\independentT}{\perp}}
\def\independentT#1#2{\mathrel{\rlap{$#1#2$}\mkern2mu{#1#2}}}
\input{tcilatex}
\numberwithin{equation}{section}
\numberwithin{assumption}{section}

\begin{document}

\title{Specification Tests for the Propensity Score}
\author{Pedro H. C. Sant'Anna\thanks{%
Department of Economics, Vanderbilt University, VU Station B \#351819, 2301
Vanderbilt Place, Nashville, TN 37235-1819, USA. Email:
pedro.h.santanna@vanderbilt.edu} 
%EndAName
 \and Xiaojun Song\thanks{%
Department of Business Statistics and Econometrics, Guanghua School of
Management and Center for Statistical Science, Peking University, Beijing,
100871, China. Email: sxj@gsm.pku.edu.cn} 
%EndAName
}
\maketitle

\begin{abstract}
\onehalfspacing{
This paper proposes new nonparametric diagnostic tools to assess the
asymptotic validity of different treatment effects estimators that rely on the correct
specification of the propensity score. We derive a particular restriction relating the propensity score 
distribution of treated and control groups, and develop specification tests based upon it. 
The resulting tests do not suffer from the
\textquotedblleft curse of dimensionality\textquotedblright\ when the vector
of covariates is high-dimensional, are fully data-driven, do not
require tuning parameters such as bandwidths, and are able to detect a broad
class of local alternatives converging to the null at the parametric rate $n^{-1/2}$, with $n$ the sample size. We show that the use of an orthogonal
projection on the tangent space of nuisance parameters facilitates the simulation of critical values by means of a multiplier
bootstrap procedure, and can lead to power gains. The finite sample performance of the tests is examined
by means of a Monte Carlo experiment and an empirical application.
Open-source software is available for implementing the proposed tests. 
}
\end{abstract}

\pagebreak

\section{Introduction}

The propensity score, which is defined as the conditional probability of
receiving treatment given covariates, is one of the most widely used tools
for causal inference. Part of its popularity can be credited to the seminal
result of \cite{Rosenbaum1983}: if the treatment assignment is independent
of the potential outcomes conditional on a vector of covariates, then one
can obtain unbiased and consistent estimators of different treatment effect
measures by adjusting for the propensity score alone, greatly reducing the
dimensionality of the underlying problem. Several methods that exploit this
important insight are now an essential part of the applied researcher's
toolkit. Examples include matching, see e.g. \cite{Rosenbaum1985}, \cite%
{Heckman1997} and \cite{Abadie2016}; inverse probability weighting (IPW),
see e.g. \cite{Rosenbaum1987a}, \cite{Hirano2003} and \cite{Donald2013};
regression methods, see e.g. \cite{Hahn1998} and \cite{Firpo2007}; and many
others. For literature reviews, see \cite{Heckman2007} and \cite{Imbens2009}.

Despite their popularity, a main concern of these methods is that the
propensity score is usually unknown, and therefore has to be estimated.
Given the high dimensionality of available covariates, researchers are
usually coerced to adopt a parametric model for the propensity score since
nonparametric estimation methods suffer from the \textquotedblleft curse of
dimensionality\textquotedblright , implying that the resulting treatment
effect estimators can have considerably poor properties, even for large
sample sizes. Such a common practice raises the important issue of model
misspecification. Indeed, as shown by \cite{Frolich2004}, \cite{Millimet2009}%
, \cite{Huber2013} and \cite{Busso2014}, propensity score misspecifications
can lead to misleading treatment effect estimates.

In this paper we propose new specification tests for parametric propensity
score models. Our proposal builds on the common practice of comparing the
density of the propensity score between treated and control groups to
determine the covariate overlap region, see e.g. \cite{Heckman1998}.
However, instead of comparing conditional densities, we focus on comparing
conditional cumulative distribution functions (CDFs). In particular, we
derive a restriction between the propensity score CDFs among treated and
control groups that gives information on overlapping\footnote{%
We thank an anonymous referee for making this suggestion.}, show that such a
restriction is equivalent to a particular infinite number of unconditional
moment conditions, and develop tests based upon it.

In contrast to existing proposals, our tests are fully data-driven, do not
require user-chosen tuning parameters such as bandwidths, and are able to
detect a broad class of local alternatives converging to the null at the
parametric rate $n^{-1/2}$, with $n$ the sample size. Furthermore, our tests
do not suffer from the \textquotedblleft curse of
dimensionality\textquotedblright\ when the vector of covariates is of high
dimensionality, and have greater power than competing tests for many
alternatives. Of course, such power gains do not come without a cost: there
exist some classes of alternative hypotheses against which our tests have
trivial power. Nonetheless, we believe that such a compromise is reasonable
since, as pointed out by \cite{Janssen2000} and \cite{Escanciano2009a},
achieving reasonable power over all possible directions seems hopeless.

The proposal closest to ours is \cite{Shaikh2009}. Despite using a similar
characterization of the null hypothesis as \cite{Shaikh2009}, our proposal
greatly differs from theirs. Whereas \cite{Shaikh2009} adopts the local
smoothing approach, see e.g. \cite{Hardle1993}, \cite{Zheng1996}, \cite%
{Fan1996} and \cite{Li1998}, we adopt the integrated conditional moment
(ICM) approach, see e.g. \cite{Bierens1982, Bierens1990}, \cite{Bierens1997}%
, \cite{Stute1997} and \cite{Escanciano2006b}. As a consequence, our
approach inherits some advantages of ICM when compared to \cite{Shaikh2009}.
First, our tests do not require delicate bandwidth choice, unlike \cite%
{Shaikh2009}'s test whose performance can be sensitive to it. Second, in
contrast with \cite{Shaikh2009}, our approach has power against local
alternatives converging to the null at the parametric rate.

Another popular procedure to assess misspecification of the propensity score
model is to use \textquotedblleft balancing\textquotedblright\ tests.
Initially proposed by \cite{Rosenbaum1985}, these tests consist of assessing
if each covariate is independent of the treatment assignment, conditional on
the propensity score. This is often implemented examining whether moments
(usually just the mean) of the observable characteristics between the two
\textquotedblleft matched\textquotedblright\ or \textquotedblleft
weighted\textquotedblright\ groups are the same; see e.g. \cite{Dehejia2002}
and \cite{Smith2005}. One should bear in mind that because \textquotedblleft
balancing\textquotedblright\ tests are usually based on a finite number of
orthogonality conditions, there are uncountably many directions of
misspecification that cannot be detected with these tests. Furthermore, as
shown by \cite{Lee2013a}, balancing tests may have size distortions due to
the \textquotedblleft multiple testing problem\textquotedblright , the
failure to account for the estimation effect of the propensity score, and
poor covariate overlap. Such drawbacks put at stake the reliability of many
of these procedures. Our proposal, on the other hand, does not suffer from
these.

Our paper also contributes to the literature on ICM tests. What appears
distinctive to our approach is that $(i)$ we exploit the dimension-reduction
coming from our derived restriction between propensity score CDFs, and $%
\left( ii\right) $ we acknowledge our lack of knowledge of the
\textquotedblleft true\textquotedblright\ correct specification of the
propensity score, by means of an orthogonal projection onto the tangent
space of nuisance parameters. The result of $(i)$ and $(ii)$ is a test with
improved power properties, and with a simple bootstrap implementation. The
power improvement due to the dimension reduction has been noticed by \cite%
{Stute2002}, \cite{Escanciano2006b} and \cite{Shaikh2009}, whereas the power
improvement due to the use of orthogonal projections has been noticed in
different contexts, see e.g. \cite{Neyman1959}, and more recently, \cite%
{Bickel2006} and \cite{Escanciano2014a}. To the best of our knowledge, our
proposal is the first to incorporate both procedures.

As mentioned above, our paper is related to the relatively scarce literature
on projection-based specification tests, see e.g. \cite{Escanciano2009simple}
and \cite{Escanciano2014a} for two notable exceptions. \cite%
{Escanciano2009simple} proposes a simple bootstrap testing procedure for
conditional moment restrictions that acknowledges specifically the fact that
the nuisance parameters are unknown and introduces the projection
methodology, while \cite{Escanciano2014a} propose a projection-based testing
procedure for linear quantile regression using a related projection weight
function. Our proposal builds on these papers, with the important difference
that our test statistics also exploit the dimension-reduction property of
the propensity score.

The rest of the paper is organized as follows. In Section \ref{framework} we
present the testing framework and derive the restriction upon which our
tests are based. The asymptotic properties of our tests are established in
Section \ref{asythe}. We next examine the finite sample properties of our
tests by means of a Monte Carlo study in Section \ref{MC}. We provide an
empirical illustration of our procedures in Section \ref{application}.
Section \ref{conclusion} concludes. Mathematical proofs are gathered in an
appendix at the end of the article.

Finally, all proposed tests discussed in this article can be implemented via
open-source R package \textit{pstest,} which is freely available from GitHub
(\url{https://github.com/pedrohcgs/pstest}).

\section{Testing Framework\label{framework}}

\subsection{Background}

Let $D$ be a binary random variable that indicates participation in the
program, i.e. $D=1$ if the individual participates in the treatment and $D=0$
otherwise. Define $Y\left( 1\right) $ and $Y\left( 0\right) $ as the
potential outcomes under treatment and control, respectively. The realized
outcome of interest is $Y=DY\left( 1\right) +\left( 1-D\right) Y\left(
0\right) $, and $X$\textbf{\ }is an observable $d\times 1$ vector of
pre-treatment covariates.\ Denote the support of $X$ by $\mathcal{X\subseteq 
}\mathbb{R}^{d}$ and the propensity score $p\left( x\right) =\mathbb{P}%
\left( D=1|X=x\right) $. We have a random sample $\left\{ \left(
Y_{i},D_{i},X_{i}^{\prime }\right) ^{\prime }\right\} _{i=1}^{n}$ of size $%
n\geq 1$ from $\left( Y,D,X^{\prime }\right) ^{\prime }$. Throughout the
rest of this article, all random variables are defined on a common
probability space $\left( \Omega ,\mathcal{A},\mathbb{P}\right) .$

The main goal in causal inference is to assess the effect of a treatment $D$
on the outcome of interest $Y$. The most popular parameters of interest
include the average treatment effect, $ATE=\mathbb{E}\left[ Y\left( 1\right)
-Y\left( 0\right) \right] $, and the average treatment effect on the
treated, $ATT=\mathbb{E}\left[ Y\left( 1\right) -Y\left( 0\right) |D=1\right]
$. Note that such parameters of interest depend on potential outcomes $%
Y\left( 1\right) $ and $Y\left( 0\right) $ which cannot be jointly observed
for the same individual, precluding estimating $ATE$ and $ATT$ using their
sample analogues. One of the most popular identification strategies in
policy evaluation that resolves such difficulty is to assume that selection
into treatment is solely based on observable characteristics, the so-called
unconfoundedness setup, see e.g. \cite{Rosenbaum1983}. Formally, the
unconfoundedness setup requires the following assumptions:

\begin{assumption}
\label{unc}$\left( Y\left( 1\right) ,Y\left( 0\right) \right) 
%TCIMACRO{\TeXButton{indep}{\independent}}%
%BeginExpansion
\independent%
%EndExpansion
D|X$.
\end{assumption}

\begin{assumption}
\label{common support}$\forall x\in \mathcal{X}$, $0<p\left( x\right) <1.$
\end{assumption}

As shown by \cite{Rosenbaum1987a}, under Assumptions \ref{unc}-\ref{common
support}, $ATE$ and $ATT$ are identified by 
\begin{equation*}
ATE=\mathbb{E}\left[ \left( \frac{D}{p\left( X\right) }-\frac{\left(
1-D\right) }{1-p\left( X\right) }\right) Y\right]\quad \text{and}\quad ATT=%
\frac{\mathbb{E}\left[ \left( D-\dfrac{p\left( X\right) \left( 1-D\right) }{%
1-p\left( X\right) }\right) Y\right] }{\mathbb{E}\left[ D\right] },
\end{equation*}%
respectively. This result motivates the two-step procedure in which one
first estimates the propensity score, computes its estimated values $\hat{p}%
\left( X_i\right) $, and then uses the analogy principle to estimate $ATE$
and $ATT$, that is,%
\begin{eqnarray*}
\widehat{ATE}_{n} &=&n^{-1}\sum_{i=1}^{n}\left[ \left( \frac{D_{i}}{\hat{p}%
\left( X_{i}\right) }-\frac{\left( 1-D_{i}\right) }{1-\hat{p}\left(
X_{i}\right) }\right) Y_{i}\right] , \\
\widehat{ATT}_{n} &=&\frac{n^{-1}\sum_{i=1}^{n}\left[ \left( D_{i}-\dfrac{%
\hat{p}\left( X_{i}\right) \left( 1-D_{i}\right) }{1-\hat{p}\left(
X_{i}\right) }\right) Y_{i}\right] }{n^{-1}\sum_{i=1}^{n}D_{i}}.
\end{eqnarray*}%
Alternatively, one could estimate $ATE$ and $ATT$ using propensity score
matching, see e.g. \cite{Rosenbaum1983}, \cite{Heckman1998a} and \cite%
{Abadie2016}.

In order to ensure that such estimators are well-defined and stable, it is
important to assess the overlap between the distribution of the propensity
score among treatment and control groups, i.e. to check whether the
propensity score is bounded away from zero and one, and if the support of
the propensity score in both groups are nearly the same, see e.g. \cite%
{Heckman1998}, \cite{Smith2005}, \cite{Crump2009} and \cite{Khan2010}.
Following \cite{Heckman1998}, it is now routine to compare kernel density
estimates of the propensity score among treated and control samples to
determine the common support region. In cases where there is strong overlap
one proceeds as described above, otherwise, one usually considers trimmed
samples, see e.g. \cite{Crump2009} and \cite{Sasaki2018}.

Although kernel density estimators are popular, they involve choosing tuning
parameters such as bandwidths and often suffer from boundary bias. Of course
such inconveniences can be easily avoided if one focuses on CDFs instead of
densities. In the following we show that propensity score overlap implies a
particular set of restrictions between the CDFs of treated and control
groups, and that these restrictions can form the basis for testing the
correct specification of propensity score models.

Assume that the propensity score $p\left( X\right) $ has a density with
respect to a dominating measure, and that the density is bounded away from
zero and infinity uniformly over its support. The following lemma builds on 
\cite{Shaikh2009} and formalizes the above discussion.

\begin{lemma}
\label{lem1}Let $\alpha =\mathbb{P}\left( D=0\right) /\mathbb{P}\left(
D=1\right) $ and assume that $0<\mathbb{P}\left( D=1\right) <1$. If $%
0<p\left( X\right) <1~a.s.,$ then%
\begin{equation}
\mathbb{E}\left[ 1\left\{ p\left( X\right) \leq u\right\} |D=1\right]
=\alpha ~\mathbb{E}\left[ \frac{p\left( X\right) }{1-p\left( X\right) }%
1\left\{ p\left( X\right) \leq u\right\} |D=0\right],~\forall u\in \left[ 0,1%
\right] .  \label{cs1}
\end{equation}%
Furthermore, (\ref{cs1}) holds if and only if 
\begin{equation}
\mathbb{E}\left[ \left( D-p\left( X\right) \right) 1\left\{ p\left( X\right)
\leq u\right\} \right] =0,~\forall u\in \left[ 0,1\right] .  \label{ort1}
\end{equation}
\end{lemma}

Lemma \ref{lem1} implies that, when the propensity score is correctly
specified, one can expect that the sample analogue of (\ref{cs1}) should
hold. Thus, (\ref{cs1}) provides a graphical diagnostic tool for propensity
score misspecification; see Lemma 3.2 of \cite{Sloczynski2018} for a result
related to (\ref{cs1}). Perhaps more importantly, note that (\ref{ort1})
provides an infinite number of simple unconditional moment restrictions that
can be used to formally test whether a parametric model for the propensity
score is correctly specified or not.

Motivated from Lemma \ref{lem1}, we seek to test whether a parametric
putative model for $p\left( x\right) $ is correctly specified based on 
\begin{equation}
H_{0}:\mathbb{E}[(D-q(X,\theta _{0}))1\{q(X,\theta _{0})\leq u\}]=0\,\,\,%
\text{for some}\,\,\,\theta _{0}\in \Theta \,\,\,\text{and for all}%
\,\,\,u\in\Pi,  \label{equiv}
\end{equation}%
where $\Theta \subset \mathbb{R}^{k}$, $\Pi=\left[ 0,1\right]$ is the unit
interval, and $q\left( X,\theta \right) :$ $\mathcal{X\times }\Theta \mapsto %
\left[ 0,1\right] $ is a family of parametric functions known up to the
finite dimensional parameter $\theta $. Common specifications for $q\left(
X,\theta \right) $ in empirical applications are the Probit, $\Phi \left(
X^{\prime }\theta \right) ,$ and the Logit, $\Lambda \left( X^{\prime
}\theta \right) $, where $\Phi \left( \cdot \right) $ and $\Lambda \left(
\cdot \right) $ are the normal and logistic link functions, respectively.

Note that (\ref{equiv})\ can be equivalently written as 
\begin{equation}
H_{0}:\mathbb{E}\left[ D-q\left( X,\theta _{0}\right) |q\left( X,\theta
_{0}\right) \right] =0~a.s.\,\,\,\text{for some}\,\,\,\theta _{0}\in \Theta ,
\label{h0}
\end{equation}%
see e.g. \cite{Stute1997}\footnote{%
Alternative representations of $H_{0}$ are also possible, see e.g. \cite%
{Bierens1997}, \cite{Escanciano2006a} and a previous version of this article.%
}. Thus, in order to assess $H_{0}$, one can either use the infinite number
of unconditional moment restrictions in (\ref{equiv}), or the conditional
moment restriction in (\ref{h0}). In this article we use (\ref{equiv})
whereas \cite{Shaikh2009} exploits (\ref{h0}). More precisely, \cite%
{Shaikh2009} consider a test statistic based on%
\begin{equation}
\hat{V}_{n}\left( h_{n}\right) =\frac{1}{n\left( n-1\right) }%
\sum_{i=1}^{n}\sum_{j=1,j\not=i}^{n}\frac{1}{h_{n}}K\left( \frac{q\left(
X_{i},\hat{\theta}_{n}\right) -q\left( X_{j},\hat{\theta}_{n}\right) }{h_{n}}%
\right) \varepsilon _{i}\left( \hat{\theta}_{n}\right) \varepsilon
_{j}\left( \hat{\theta}_{n}\right) ,  \label{shaikh}
\end{equation}%
where $\varepsilon _{i}\left( \hat{\theta}_{n}\right) =D_{i}-q\left( X_{i},%
\hat{\theta}_{n}\right) $, $\hat{\theta}_{n}$ is a $\sqrt{n}-$consistent
estimator of $\theta _{0}$ under $H_{0}$, $h_{n}$ is a positive scalar
bandwidth parameter converging to zero at a suitable rate as $n\rightarrow
\infty $, and $K\left( \cdot \right) $ is a kernel function. Note that, in
addition to the estimation of $\theta _{0}$ under $H_{0}$, \cite{Shaikh2009}%
's procedure requires local smoothing of the data, implying that its finite
sample properties rely on the adequate choice of the smoothing parameter $%
h_{n}$, a task that is far from trivial in testing problems. Since our
approach is based on (\ref{equiv}) and only involves unconditional
expectations, our testing procedure is free of tuning parameters such as
bandwidth sequence $h_{n}$.

Tests based on a continuum of unconditional moment restrictions such as (\ref%
{equiv}) fall into the ICM approach, see \cite{Gonzalez-Manteiga2013} for a
review. Nonetheless, our tests have two main differences with respect to the
standard ICM tests. First, (\ref{equiv}) depends on $X$ only through the
propensity score model under $H_{0}$, a one-dimensional (though unknown)
function. As a consequence, the ICM in (\ref{equiv}) is insensitive to the
dimension $d$ of the explanatory variables $X$, avoiding the so-called
\textquotedblleft curse of dimensionality\textquotedblright . Second, in
contrast to the standard ICM tests, we explicitly acknowledge that $\theta
_{0}$ is a nuisance parameter in testing (\ref{equiv}) by proposing to use
orthogonal projections on the tangent space of nuisance parameters. As
discussed in the Introduction, the use of orthogonal projections leads to
important advantages. In the next subsection we describe how we construct
such projection-based tests, paying particular attention to the role played
by the orthogonal projection; see also \cite{Escanciano2009simple} and \cite%
{Escanciano2014a} for related results in different contexts.

\begin{Rem}
\label{rem:unc}Although the identification of treatment effect parameters
such as the $ATE$ relies on both Assumptions \ref{unc} and \ref{common support}%
, the results in Lemma \ref{lem1} (and the null hypothesis (\ref{equiv})) do
not involve outcome data and are therefore well motivated even when
Assumption \ref{unc} may not hold. This situation may arise in decomposition
exercises, see e.g. \cite{Fortin2011} for a review of decomposition methods
in economics. With respect to Assumption (\ref{common support}), we note that it allows
propensity scores to be arbitrarily close to zero and one and hence it is
not very restrictive. In fact, given that the unconditional moment condition in (\ref{ort1}%
) does not involve random denominators, weak covariate overlap does not play
a major role in our testing procedure. However, weak covariate overlap may
lead to irregular treatment effect estimators, see e.g. \cite{Khan2010}.
\end{Rem}

\subsection{Projection-based specification tests\label{prop-test}}

Recall that $\varepsilon _{i}\left( \theta \right) =D_{i}-q\left(
X_{i},\theta \right) .$ For all $u\in \Pi $, define 
\begin{equation}
\mathcal{P}_{n}1\left\{ q(X,\theta )\leq u\right\} =1\left\{ q(X,\theta
)\leq u\right\} -g^{\prime }(X,\theta )\Delta _{n}^{-1}\left( \theta \right)
G_{n}\left( u,\theta \right) ,  \label{project}
\end{equation}%
where $g(x,\theta )=\partial q(x,\theta )/\partial \theta$ is the score
function of $q(x,\theta )$, 
\begin{equation*}
G_{n}(u,\theta )=\frac{1}{n}\sum_{i=1}^{n}g(X_{i},\theta )1\left\{
q(X_{i},\theta )\leq u\right\} ,
\end{equation*}%
and 
\begin{equation*}
\Delta _{n}\left( \theta \right) =\frac{1}{n}\sum_{i=1}^{n}g(X_{i},\theta
)g^{\prime }(X_{i},\theta ).
\end{equation*}

Given a random sample $\left\{ \left( D_{i},X_{i}^{\prime }\right) ^{\prime
}\right\} _{i=1}^{n}$, our test statistics are based on continuous
functionals of the projection-based empirical process $\hat{R}_{n}^{p}(u)$, 
\begin{equation}
\hat{R}_{n}^{p}(u)\equiv \frac{1}{\sqrt{n}}\sum_{i=1}^{n}\varepsilon _{i}(%
\hat{\theta}_{n})\mathcal{P}_{n}1\left\{ q(X_{i},\hat{\theta}_{n})\leq
u\right\} ,  \label{emp-process}
\end{equation}%
where $\hat{\theta}_{n}$ is a $\sqrt{n}-$consistent estimator for $\theta
_{0}$ under $H_{0}$. Two popular examples of such functionals are the Cram%
\'{e}r-von Mises-type and Kolmogorov-Smirnov-type functionals, 
\begin{align}
CvM_{n}& =\int_{\Pi }\left\vert \hat{R}_{n}^{p}(u)\right\vert
^{2}\,F_{n}(du)=\frac{1}{n}\sum_{i=1}^{n}\left[ \hat{R}_{n}^{p}\left(
q\left( X_{i},\hat{\theta}_{n}\right) \right) \right] ^{2},  \label{cvm} \\
KS_{n}& =\sup_{u\in \Pi }\left\vert \hat{R}_{n}^{p}(u)\right\vert ,
\label{ks}
\end{align}%
respectively, where $F_{n}(u)=n^{-1}\sum_{i=1}^{n}1\left( q\left( X_{i},\hat{%
\theta}_{n}\right) \leq u\right) $ is the empirical distribution function
(EDF) of $q\left( X_{i},\hat{\theta}_{n}\right) $, $1\leq i\leq n.$

At this point, one may wonder why our test statistics are based on the
empirical process $\hat{R}_{n}^{p}(u)$ (\ref{emp-process}) instead of the
usual sample analogue of (\ref{equiv}), 
\begin{equation}
\hat{R}_{n}(u)\equiv \frac{1}{\sqrt{n}}\sum_{i=1}^{n}\varepsilon _{i}(\hat{%
\theta}_{n})1\left\{ q(X_{i},\hat{\theta}_{n})\leq u)\right\} ,  \label{R1}
\end{equation}%
i.e. the \textquotedblleft unprojected\textquotedblright\ analogue of $\hat{R%
}_{n}^{p}(u)$. To answer such a query, note that under $H_{0}$ and some weak
regularity conditions given in Section \ref{null-sec}, the unprojected
process $\hat{R}_{n}(u)$ can be decomposed as%
\begin{multline}
\hat{R}_{n}(u)=\frac{1}{\sqrt{n}}\sum_{i=1}^{n}\varepsilon _{i}(\theta
_{0})1\left\{ q(X_{i},\theta _{0})\leq u)\right\} - \\
\sqrt{n}(\hat{\theta}_{n}-\theta _{0})^{\prime }\mathbb{E}\left[ g(X,\theta
_{0})1\left\{ q(X,\theta _{0})\leq u)\right\} \right] +o_{p}\left( 1\right) ,
\label{linrepR}
\end{multline}%
uniformly in $u\in \Pi $, see Lemma \ref{lemmaA3} in the Appendix. The
asymptotic representation in (\ref{linrepR}) implies that the effect of
replacing $\theta _{0}$ by $\hat{\theta}_{n}$ is non-negligible, and
therefore the asymptotic null distributions of tests based on (\ref{R1}) are
sensitive to the estimator $\hat{\theta}_{n}$ being used. As a consequence,
for a given parametric specification $p\left( x\right) =q(X,\theta _{0})$,
the asymptotic null distributions of tests based on (\ref{R1}) will depend
on whether one estimates $\theta _{0}$ using maximum likelihood (ML),
nonlinear least squares (NLS), or generalized method of moments (GMM), even
though the underlying specification for the propensity score is the same
across these estimation methods.

The projection-based process $\hat{R}_{n}^{p}(u)$, on the other hand, avoids
such drawback since 
\begin{equation}
\mathbb{E}\left[ g(X,\theta _{0})\mathcal{P}1\left\{ q(X,\theta _{0})\leq
u)\right\} \right]\equiv 0  \label{mort}
\end{equation}%
almost everywhere in $u\in \Pi $, where 
\begin{equation}
\mathcal{P}1\left\{ q(X,\theta )\leq u)\right\} = 1\left\{ q(X,\theta )\leq
u)\right\} -g^{\prime }(X,\theta )\Delta ^{-1}\left( \theta \right) G\left(
u,\theta \right) ,  \label{projection}
\end{equation}%
with 
\begin{equation*}
G(u,\theta )=\mathbb{E}\left[ g(X,\theta )1\left\{ q(X,\theta )\leq
u\right\} \right] ,
\end{equation*}%
and 
\begin{equation*}
\Delta \left( \theta \right) =\mathbb{E}\left[ g(X,\theta )g^{\prime
}(X,\theta )\right] .
\end{equation*}%
The intuition behind (\ref{mort}) is simple. First, note that $\Delta
^{-1}\left( \theta \right) G\left( u,\theta \right) $ is the vector of
linear projection coefficients of regressing $1\left\{ q(X,\theta )\leq
u\right\} $ on $g(X,\theta )$. Thus, it follows that $g(X,\theta )^{\prime
}\Delta ^{-1}\left( \theta \right) G\left( u,\theta \right) $ is the best
linear predictor of $1\left\{ q(X,\theta )\leq u\right\} $ given $g(X,\theta
)$, and that (\ref{projection}) is nothing more than the associated
projection error, which is by definition orthogonal to $g(X,\theta )$. As a
consequence of (\ref{mort}), it follows that under some weak regularity
conditions, uniformly in $u\in \Pi $, 
\begin{equation*}
\hat{R}_{n}^{p}(u)=R_{n0}^{p}(u)+o_{p}\left( 1\right) ,
\end{equation*}%
where 
\begin{equation}
R_{n0}^{p}(u)=\frac{1}{\sqrt{n}}\sum_{i=1}^{n}\varepsilon _{i}(\theta _{0})%
\mathcal{P}1\left\{ q(X_{i},\theta _{0})\leq u)\right\} ,  \label{rn0}
\end{equation}%
see Theorem \ref{thh0} in Section \ref{null-sec}. Thus, $\hat{R}_{n}^{p}(u)$
is (asymptotically) invariant to the choice of estimator $\hat{\theta}_{n}$.
Furthermore, as we discuss in Section \ref{boots}, the above asymptotic
representation of $\hat{R}_{n}^{p}(u)$ in terms of $R_{n0}^{p}(u)$ allows
for a multiplier-type bootstrap procedure that greatly simplifies the
computation of asymptotically valid critical values.

In summary, by focusing on the projection-based process $\hat{R}_{n}^{p}(u)$
instead of the more traditional process $\hat{R}_{n}(u),$ our proposed test
statistics are $\left( a\right) $ invariant to the choice of estimator $\hat{%
\theta}_{n}$, and $\left( b\right) $ allow for simplified bootstrap
implementation. In addition, tests based on $\hat{R}_{n}^{p}(u)$ acknowledge
that deviations in the direction of the score function $g(x,\theta)$ cannot
be distinguished from deviations within the parametric model, and therefore
do not \textquotedblleft waste\textquotedblright\ power in such directions.
As a result, tests based on $\hat{R}_{n}^{p}(u$) can have higher power when
compared to tests based on $\hat{R}_{n}(u)$, though in general none of them
is strictly better than the other uniformly over the space of alternatives.
We defer the discussion of these power properties to Section \ref{powersec}.

\section{Asymptotic theory\label{asythe}}

In this section, we establish the asymptotic behavior of the
projection-based empirical process $\hat{R}_{n}^{p}(u)$ under the null
hypothesis $H_{0}$, under the fixed alternative hypothesis $H_{1}$, which is
the negation of (\ref{equiv}), and under a sequence of local alternatives $%
H_{1n}$ that converges to $H_{0}$ at the parametric rate $n^{-1/2}$, $n$
being the sample size. We also characterize classes of alternative
hypotheses against which our tests have no power, and argue that such
classes are rather exceptional. Finally, we show that critical values can be
computed with the assistance of a multiplier-type bootstrap that is easy to
implement.

\subsection{Asymptotic null distribution\label{null-sec}}

The asymptotic null distributions of our tests are the limiting
distributions of continuous functionals of $\hat{R}_{n}^{p}(u)$ under $H_{0}$%
. To derive the asymptotic results, we adopt the following notation. For a
generic set $\mathcal{G}$, let $l^{\infty }\left( \mathcal{G}\right) $ be
the Banach space of all uniformly bounded real functions on $\mathcal{G}$,
equipped with the uniform metric $\left\Vert f\right\Vert _{\mathcal{G}%
}\equiv \sup_{z\in \mathcal{G}}\left\vert f\left( z\right) \right\vert $. We
study the weak convergence of $\hat{R}_{n}^{p}(u)$ and its related processes
as elements of $l^{\infty }\left( \Pi \right) $, where $\Pi \equiv \left[ 0,1%
\right] $. Let ``$\Rightarrow $'' denote weak convergence on $\left(
l^{\infty }\left( \Pi \right) ,\mathcal{B}_{\infty }\right) $ in the sense
of J. Hoffmann-J$\phi $rgensen, where $\mathcal{B}_{\infty }$ denotes the
corresponding Borel $\sigma $-algebra - see e.g. Definition 1.3.3 in \cite%
{VanderVaart1996}.

We assume the following regularity conditions. Let $\Theta _{0}$ be an
arbitrarily small neighborhood around $\theta _{0}$ such that $\Theta
_{0}\subset \Theta $. For any $d_1\times d_2$ matrix $A=(a_{ij})$, let $%
||A|| $ denote its Euclidean norm, i.e. $||A||=[\text{tr}(AA^{\prime})] ^{1/2}$.

\begin{assumption}
\label{ass1}$(i)$ The parameter space $\Theta $ is a compact subset of $%
\mathbb{R}^{k};$ $\left( ii\right) $ the true parameter $\theta _{0}$
belongs to the interior of $\Theta $; and $\left( iii\right) $ $\left\Vert 
\hat{\theta}_{n}-\theta _{0}\right\Vert =O_{p}(n^{-1/2}).$
\end{assumption}

\begin{assumption}
\label{ass2} The parametric propensity score function $q(x,\theta )$ is
twice continuously differentiable in $\Theta _{0}$ for each $x\in\mathcal{X}$%
, with its first derivative $g(x,\theta )=\partial q(x,\theta )/\partial
\theta =(g_{1}(x,\theta ),\ldots ,g_{k}(x,\theta ))^{\prime }$ satisfying $%
\mathbb{E}[\sup_{\theta \in \Theta _{0}}||g(X,\theta )||]<\infty $ and its
second derivative satisfying $\mathbb{E}[\sup_{\theta \in \Theta
_{0}}||\partial g(X,\theta )/\partial \theta ||]<\infty $. Furthermore, the
matrix $\Delta (\theta )\equiv \mathbb{E}[g(X,\theta )g^{\prime }(X,\theta
)] $ is nonsingular in $\Theta _{0}$.
\end{assumption}

\begin{assumption}
\label{ass3}The function $F_{\theta }(u)=\mathbb{P}(q(X,\theta )\leq u)$
satisfies $\sup_{u\in \Pi }|F_{\theta _{1}}(u)-F_{\theta _{2}}(u)|\leq
C||\theta _{1}-\theta_{2}||$, where $C$ is a bounded positive number, not
depending on $\theta_1$ and $\theta_2$.
\end{assumption}

Assumptions \ref{ass1}-\ref{ass3} are weaker than related conditions in the
literature. For instance, Assumption \ref{ass1} only requires $\sqrt{n}%
\left( \hat{\theta}_{n}-\theta _{0}\right) =O_{p}\left( 1\right) ,$ but does
not require $\sqrt{n}\left( \hat{\theta}_{n}-\theta _{0}\right) $ to admit
an asymptotically linear representation. Assumption \ref{ass2} is a
condition concerning the degree of smoothness of the propensity score $%
q(x,\theta )$, and is satisfied for standard parametric models such as the
Probit and the Logit specifications. It also only requires finite first
moment of $g(X,\theta )$, instead of more than four moments as in \cite%
{Shaikh2009}. Assumption \ref{ass3} simply imposes a Lipschitz type
continuity condition on the CDF of the parametric propensity score.

\begin{Rem}
Assumption \ref{ass3} is used to prove that the class of functions $\mathcal{%
F}=\{x\mapsto 1\left\{ q(x,\theta )\leq u\right\} :u\in \Pi ,\,\theta \in
\Theta \}$ is Donsker, see Lemma \ref{lemDonsker} in the Appendix. Such
assumption is similar to condition (5.8) in \cite{Lee2011e}. Alternatively,
if $\mathcal{F}_{\theta }=\{x\mapsto q(x,\theta ):\theta \in \Theta \}$ is a
VC class of functions, the aforementioned Donsker result also follows even
without Assumption \ref{ass3}, see e.g. Example 2.1 in \cite{VanderVaart2007}%
.
\end{Rem}

Next, we derive the asymptotic behavior of the projection-based empirical
process $\hat{R}_{n}^{p}(u)$ under $H_{0}$. We do this in two steps. First,
we show that, under $H_{0}$, $\hat{R}_{n}^{p}(u)$ is asymptotically
equivalent, with respect to the supremum norm on $\Pi $, to the process $%
R_{n0}^{p}(u)$ given in (\ref{rn0}). From this result it follows that the
weak convergence under $H_{0}$ of the process $\hat{R}_{n}^{p}(u)$ can be
conveniently established from that of $R_{n0}^{p}(u)$. More importantly, the
limiting null behavior of $\hat{R}_{n}^{p}(u)$ does not depend on $\hat{%
\theta}_{n}$ nor how $\hat\theta_n$ is obtained.

\begin{theorem}
\label{thh0}Let Assumptions \ref{ass1}-\ref{ass3} hold. Then, under $H_{0}$,
we have that 
\begin{equation*}
\sup_{u\in \Pi }\left\vert \hat{R}_{n}^{p}(u)-R_{n0}^{p}(u)\right\vert
=o_{p}(1),
\end{equation*}%
and 
\begin{equation*}
\hat{R}_{n}^{p}(u)\Rightarrow R_{\infty }^{p}\text{,}
\end{equation*}%
where $R_{\infty }^{p}$ denotes a Gaussian process with mean zero and
covariance structure given by%
\begin{equation}
K^{p}(u_{1},u_{2})=\mathbb{E}\left[q(X,\theta _{0})\left(1-q(X,\theta
_{0})\right)\mathcal{P}1\left\{q(X,\theta _{0})\leq u_{1}\right\}\mathcal{P}%
1\left\{q(X,\theta _{0})\leq u_{2}\right\}\right].  \label{kw}
\end{equation}
\end{theorem}

Theorem \ref{thh0} and the continuous mapping theorem (CMT), see e.g.
Theorem 1.3.6 in \cite{VanderVaart1996}, yield the asymptotic null
distributions of continuous functionals of $\hat{R}_{n}^{p}(u),$ including
the test statistics $CvM_{n}$ and $KS_{n}$ given in (\ref{cvm}) and (\ref{ks}%
), respectively.

\begin{corollary}
\label{corho}Under the assumptions of Theorem \ref{thh0} and $H_{0}$, for
any continuous functional $\Gamma (\cdot )$ from $l^{\infty }\left(
\Pi\right) $ to $\mathbb{R}$, we have 
\begin{equation*}
\Gamma (\hat{R}_{n}^{p})\xrightarrow{d}\Gamma (R_{\infty }^{p}).
\end{equation*}%
Furthermore, 
\begin{equation*}
CvM_{n}\xrightarrow{d}CvM_{\infty }:=\int_{\Pi }\left\vert R_{\infty
}^{p}(u)\right\vert ^{2}\,dF_{\theta _{0}}(u),
\end{equation*}%
where $F_{\theta _{0}}(u)=\mathbb{P}\left( q\left( X,\theta _{0}\right) \leq
u\right) $ denotes the cumulative distribution function of $q\left( X,\theta
_{0}\right)$, and 
\begin{equation*}
KS_{n}\xrightarrow{d}KS_{\infty }:=\sup_{u\in \Pi }\left\vert R_{\infty
}^{p}(u)\right\vert .
\end{equation*}
\end{corollary}

Note that the integrating measure in $CvM_{n}$ is a random measure, but
Corollary \ref{corho} shows that the asymptotic distribution is not affected
by this fact. Further details can be found in the Appendix A.

\subsection{Asymptotic power \label{powersec}}

Now, we investigate the power properties of tests based on continuous
functionals $\Gamma (\hat{R}_{n}^{p})$, like $CvM_{n}$ and $KS_{n}$ in (\ref%
{cvm}) and (\ref{ks}), respectively. We consider fixed alternatives, and a
sequence of local alternatives $H_{1n}$ that converges to $H_{0}$ at the
parametric rate $n^{-1/2}$.

\subsubsection{Power against fixed alternatives}

Next theorem analyzes the asymptotic properties of our tests under fixed
alternatives of the type%
\begin{equation}
H_{1}:\mathbb{E}[(D-q(X,\theta ))1\{q(X,\theta )\leq u\}]\neq 0\,\,\,\text{%
for all}\,\,\,\theta \in \Theta \,\,\,\text{and for some}\,\,\,u\in \Pi,
\label{h1}
\end{equation}%
where $\Pi =[0,1]$ is the unit interval. Note that $H_{1}$ is simply the
negation of $H_{0}$ in (\ref{equiv}).

\begin{theorem}
\label{thh1} Suppose Assumptions \ref{ass1}-\ref{ass3} hold. Then, under the
fixed alternative hypothesis $H_{1}$ in (\ref{h1}), we have that 
\begin{equation*}
\sup_{u\in \Pi }\left\vert \frac{1}{\sqrt{n}}\hat{R}_{n}^{p}(u)-\mathbb{E}%
\left[ \left( p\left( X\right) -q\left( X,\theta _{0}\right) \right) 
\mathcal{P}1\left\{ q\left( X,\theta _{0}\right) \leq u\right\} \right]
\right\vert =o_{p}\left( 1\right) .
\end{equation*}
\end{theorem}

From Theorem \ref{thh1}, we see that test statistics of the form of $\Gamma (%
\hat{R}_{n}^{p})$ are not consistent against all fixed alternative
hypotheses in (\ref{h1}), but only those not collinear to the score function 
$g(X,\theta_0)$. To see this, note that 
\begin{multline*}
\mathbb{E}\left[ \left( p\left( X\right) -q\left( X,\theta _{0}\right)
\right) \mathcal{P}1\left\{ q\left( X,\theta _{0}\right) \leq u\right\} %
\right] = \\
\mathbb{E}\left[ \left( p\left( X\right) -q\left( X,\theta _{0}\right)
\right) 1\left\{ q\left( X,\theta _{0}\right) \leq u\right\} \right] - \\
\mathbb{E}\left[ \left( p\left( X\right) -q\left( X,\theta _{0}\right)
\right) g^{\prime }(X,\theta _{0})\right] \Delta ^{-1}\left( \theta
_{0}\right) G\left( u,\theta _{0}\right)
\end{multline*}%
is equal to zero under (\ref{h1}) if $p\left( X\right) -q\left( X,\theta
_{0}\right)$ and $g(X,\theta _{0})$ are collinear almost surely. We do not
see this as a limitation. First, when one estimates $\theta _{0}$ using the
NLS method, the population first order condition for $\theta _{0}$ sets $%
\mathbb{E}\left[ \left( D-q\left( X,\theta _{0}\right) \right) g^{\prime
}(X,\theta _{0})\right] =0,$ implying that, for some $u\in \Pi ,$ 
\begin{equation*}
\mathbb{E}\left[ \left( p\left( X\right) -q\left( X,\theta _{0}\right)
\right) \mathcal{P}1\left\{ q\left( X,\theta _{0}\right) \leq u\right\} %
\right] =\mathbb{E}\left[ \left( p\left( X\right) -q\left( X,\theta
_{0}\right) \right) 1\left\{ q\left( X,\theta _{0}\right) \leq u\right\} %
\right]\neq 0.
\end{equation*}%
As a consequence, our projection-based tests would be consistent against all
alternative hypotheses of the type of (\ref{h1}), avoiding the
aforementioned problem.

Second, and perhaps more importantly, even when one does not use NLS to
estimate $\theta _{0},$ we argue that the lack of power against alternatives
collinear to the score function $g(X,\theta _{0})$ is not a main concern. As
shown by \cite{Escanciano2009a}, every test based on ICM approach has
trivial local power against these alternatives, and as a consequence, the
global power of all ICM tests in the direction of the score function will
also be low, see e.g. \cite{Strasser1990}. In fact, instead of considering
this as a limitation one may consider this property as a feature of our
tests: by acknowledging that such alternatives cannot be powerfully
detected, our projection-based test statistics do not waste power in such
directions, and therefore may have higher power against other, perhaps more
important, alternatives; see Section \ref{MC} for an illustration.

The above discussion raises the question: are there other classes of fixed
alternative hypotheses that our specification tests are not able to detect?
As first pointed out by \cite{Shaikh2009}, the answer is yes. Because our
test statistics depend on $X$ only through the propensity score $q(X,\theta
_{0})$, our tests will have trivial power against the class of misspecified
propensity scores, where 
\begin{equation*}
\mathbb{E}\left[ D-q\left( X,\theta _{0}\right) |X\right] =p\left( X\right)
-q\left( X,\theta _{0}\right) \not=0~
\end{equation*}%
in a set of positive probability, but%
\begin{equation*}
\mathbb{E}\left[ D-q\left( X,\theta _{0}\right) |q\left( X,\theta
_{0}\right) \right] =0\,\,\,a.s.,
\end{equation*}%
where $\theta _{0}$ now is the probability limit of $\hat{\theta}_{n}$. In
other words, tests based on (\ref{equiv}) (or equivalently on (\ref{h0})),
will have trivial power against alternatives such that $p\left( X\right) $
is different from $q\left( X,\theta _{0}\right) $ but 
\begin{equation}
\mathbb{E}\left[ p\left( X\right) |q\left( X,\theta _{0}\right) \right]
=q\left( X,\theta _{0}\right) \,\,\,a.s..  \label{weird}
\end{equation}%
The leading case of such a class of alternatives is when the propensity
score is correctly specified for a subvector of $X$, but not for the entire
vector $X$, see e.g. \cite{Shaikh2009}. Given the nonlinear nature of $%
p\left( X\right) $, one may consider such a class of alternatives rather
exceptional. However, they can still arise in practice under some particular
circumstances as we will illustrate below.

Consider the case where $X=\left( X_{1},X_{2}\right) $ and that both
covariates are relevant for the propensity score, i.e., $p\left( X\right)
=p\left( X_{1},X_{2}\right) $. Suppose that a researcher considers a Probit
model for the $p\left( \cdot \right) $ but only included $X_{1}$ as a
covariate, i.e., the researcher assume the model $q\left( X,\theta
_{0}\right) =\Phi \left( \theta _{00}+\theta _{01}X_{1}\right) $, with $\Phi
\left( \cdot \right) $ the normal link function. Of course, $p\left(
X\right) \not=q\left( X,\theta _{0}\right) $ since $q\left( X,\theta
_{0}\right) $ only includes a subset of relevant covariates. Thus, in light
of (\ref{weird}), our tests will have no power if there exists $\theta
_{0}=\left( \theta _{00},\theta _{01}\right)^{\prime }\in \Theta \subset%
\mathbb{R}^2$ such that $\mathbb{E}\left[ p\left( X_{1},X_{2}\right) |X_{1}%
\right] =\Phi \left( \theta _{00}+\theta _{01}X_{1}\right) $ $a.s.$. The
existence of such $\theta _{0}$ depends on the underlying conditional
distribution of $X_{2}$ given $X_{1}$, and also on the form of true unknown
propensity score $p\left( X_{1},X_{2}\right) $. For instance, assume that
the true propensity score is a Probit, e.g., $p\left( X_{1},X_{2}\right)
=\Phi \left( X_{1}+X_{2}\right) $, and that $X_{2}$ given $X_{1}$ follows a
normal distribution with conditional mean $\mu \left( X_{1}\right) $ and
conditional variance $\sigma ^{2}\left( X_{1}\right) $. In this case, we can
show that 
\begin{equation*}
\mathbb{E}\left[ \left. \Phi \left( X_{1}+X_{2}\right) \right\vert X_{1}%
\right] =\Phi \left( \frac{X_{1}+\mu \left( X_{1}\right) }{\sqrt{1+\sigma^2
\left( X_{1}\right)} }\right) .
\end{equation*}%
Thus, if $\mu \left( X_{1}\right) =a+bX_{1}$ and $\sigma^2 \left(
X_{1}\right) =c$ for some constants $a,b$ and $c$ with $b\neq -1$ and $c>0$,
we have that%
\begin{equation*}
\mathbb{E}\left[ \left. \Phi \left( X_{1}+X_{2}\right) \right\vert X_{1}%
\right] =\Phi \left( \frac{X_{1}+a+bX_{1}}{\sqrt{1+c}}\right) =\Phi \left( 
\frac{a}{\sqrt{1+c}}+\frac{1+b}{\sqrt{1+c}}X_{1}\right).
\end{equation*}%
Thus, in this particular case such $\theta _{0}$ does exist with $%
\theta_0\equiv\left(a/\sqrt{1+c},(1+b)/\sqrt{1+c}\right)^{\prime }$ and our
tests would have trivial power. On the other hand, if $\mu \left(
X_{1}\right) $ is nonlinear in $X_{1}$ and $\sigma ^{2}\left( X_{1}\right) $
is a nontrivial function of $X_{1}$, no such $\theta _{0}$ exists and
therefore (\ref{weird}) is ruled out. Thus, it is clear that the conditional
distribution of $X_{2}$ given $X_{1}$ plays an important role in the
\textquotedblleft empirical relevance\textquotedblright\ of alternative
hypotheses like (\ref{weird}).

The role played by the functional form of the true propensity score in the
\textquotedblleft empirical relevance\textquotedblright\ of (\ref{weird})
can be illustrated by assuming that the true propensity score is a Logit
instead of a Probit, e.g., $p\left( X_{1},X_{2}\right) =\Lambda \left(
X_{1}+X_{2}\right) $, with $\Lambda(\cdot)$ the logistic link function. In
this case, because of the nonlinear nature of $\Lambda $, there exists no $%
\theta _{0}\in \Theta $ such that $\mathbb{E}\left[ \Lambda \left(
X_{1}+X_{2}\right) |X_{1}\right] =\Phi \left( \theta _{01}+\theta
_{01}X_{1}\right) $ $a.s.$, ruling out (\ref{weird})\footnote{%
In practice, however, the power against this particular alternative can be
low since the Logit and Probit specifications are relatively
\textquotedblleft close\textquotedblright\ to each other.}.

In summary, the above discussion shows that our projection-based tests are
consistent against a broad range of alternatives, though not all. This is
the main drawback of our proposal when compared to the standard ICM
specification tests. However, in our simulations that follow we show that,
for the alternatives considered, our projection-based tests are the best or
comparable to the best tests in terms of power in finite samples. Thus, from
a practical point of view, the benefits of using our procedure can outweigh
the costs in many relevant situations.

\subsubsection{Power against local alternatives}

Next, we study the performance of our projection-based tests under a
sequence of local alternative hypotheses converging to the null at the
parametric rate $n^{-1/2}$ given by 
\begin{equation}
H_{1n}:E\left[ D-q(X,\theta _{0})|q(X,\theta _{0})\right] =\frac{r\left(
q(X,\theta _{0})\right) }{\sqrt{n}}\quad a.s.  \label{h1n}
\end{equation}%
for some $\theta _{0}\in \Theta $, where $r\left(q(X,\theta _{0})\right)$
represents directions of departure from $H_0$, and $n^{-1/2}$ indicates the
rate of convergence of $H_{1n}$ to $H_0$. The function $r:\left[ 0,1\right]
\rightarrow \mathbb{R}$ is required to satisfy the following assumption.

\begin{assumption}
\label{ass4} The function $r(q)$ is continuous in $q$ and satisfies $\mathbb{%
E}|r(q(X,\theta _{0}))|<\infty $.
\end{assumption}

\begin{theorem}
\label{thh1n} Suppose Assumptions \ref{ass1} -\ref{ass4} hold. Then, under
the local alternatives $H_{1n}$ given by (\ref{h1n}), we have 
\begin{equation*}
\hat{R}_{n}^{p}(u)\Rightarrow R_{\infty }^{p}+\Delta _{r},
\end{equation*}%
where $R_{\infty }^{p}$ is the same Gaussian process as defined in Theorem %
\ref{thh0}, and $\Delta _{r}$ is a deterministic shift function given by 
\begin{equation*}
\Delta _{r}(u)\equiv \mathbb{E}\left[r\left( q\left( X,\theta _{0}\right)
\right) \mathcal{P}1\left\{ q(X,\theta _{0})\leq u\right\} \right].
\end{equation*}
\end{theorem}

Note that, in general, the deterministic shift function $\Delta _{r}\left(
u\right) \not=0$ for at least some $u\in \Pi $, implying that tests based on
continuous even functionals of $\hat{R}_{n}^{p}(\cdot )$ will have
non-trivial power against local alternatives of the form in (\ref{h1n}). A
situation in which our tests will have trivial local power against such
alternatives is when directions $r(q\left(x,\theta _{0}\right) )$ are a
linear combination of score function $g(x,\theta _{0})$, i.e. $r(q(x,\theta
_{0}))=\beta g(x,\theta _{0})$ for some $\beta $. In such a case, the
limiting distribution of $\hat R_n^p(u)$ under $H_0$ and $H_{1n}$ is the
same so that $H_{1n}$ cannot be detected. On the other hand, note that tests
based on the local smoothing approach such as \cite{Shaikh2009} are not able
to detect alternatives of the form (\ref{h1n}).

\subsection{Computation of critical values\label{boots}}

From the above theorems, we see that the asymptotic distribution of
continuous functionals $\Gamma \left( \hat{R}_{n}^{p}\right) $ depend on the
underlying data generating process and of course on $\Gamma \left( \cdot
\right) $ itself. Furthermore, the complicated covariance structure of $%
K^{p}(\cdot ,\cdot )$ given in (\ref{kw}) does not allow for a simple
representation of $R_{\infty }^{p}$ in terms of a well-known
distribution-free Gaussian process for which critical values are readily
available. To overcome this problem, we propose to compute critical values
with the assistance of a multiplier bootstrap. The proposed procedure has
good theoretical and empirical properties, is computationally easy to
implement, and does not require computing new parameter estimates at each
bootstrap replication.

More precisely, in order to estimate the critical values, we propose to
approximate the asymptotic behavior of $\hat{R}_{n}^{p}(u)$ by that of 
\begin{equation}
\hat{R}_{n}^{p\ast }(u)\equiv \frac{1}{\sqrt{n}}\sum_{i=1}^{n}\varepsilon
_{i}(\hat{\theta}_{n})\mathcal{P}_{n}1\left\{ q\left( X_{i},\hat{\theta}%
_{n}\right) \leq u\right\} V_{i},  \label{boot}
\end{equation}%
where $\{V_{i}\}_{i=1}^{n}$ is a sequence of $i.i.d.$ random variables with
zero mean, unit variance and bounded support, independent of the original
sample $\{(D_{i},X_{i}^{\prime })^{\prime }\}_{i=1}^{n}$. A popular example
involves $i.i.d.$ Bernoulli variates $\left\{ V_{i}\right\} $ with $\mathbb{P%
}\left( V=1-\kappa \right) =\kappa /\sqrt{5}$ and $\mathbb{P}\left( V=\kappa
\right) =1-\kappa /\sqrt{5}$, where $\kappa =\left( \sqrt{5}+1\right) /2,$
as suggested by \cite{Mammen1993}.

With $\hat{R}_{n}^{p\ast }(u)$ at hands, the bootstrapped version of our
test statistics $\Gamma \left( \hat{R}_{n}^{p}\right) $ is simply given by $%
\Gamma \left( \hat{R}_{n}^{p,\ast }\right) $. For instance, the bootstrapped
versions of $CvM_{n}$ and $KS_{n}$ in (\ref{cvm}) and (\ref{ks}),
respectively, are given by%
\begin{align*}
CvM_{n}^{\ast }& =\frac{1}{n}\sum_{i=1}^{n}\left[ \hat{R}_{n}^{p\ast }\left(
q\left( X_{i},\hat{\theta}_{n}\right) \right) \right] ^{2}, \\
KS_{n}^{\ast }& =\sup_{u\in \Pi }\left\vert \hat{R}_{n}^{p\ast
}(u)\right\vert .
\end{align*}%
The asymptotic critical values are then estimated by%
\begin{equation*}
c_{n,\alpha }^{^{\Gamma ~\ast }}\equiv \inf \left\{ c_{\alpha }\in \lbrack
0,\infty ):\lim_{n\rightarrow \infty }\mathbb{P}_{n}^{^{\ast }}\left\{
\Gamma \left( \hat{R}_{n}^{p,\ast }\right) >c_{\alpha }\right\} =\alpha 
\text{ }\right\} ,
\end{equation*}%
where $\mathbb{P}_{n}^{^{\ast }}$ means bootstrap probability, i.e.
conditional on the original sample $\{(D_{i},X_{i}^{\prime })^{\prime
}\}_{i=1}^{n}$. In practice, $c_{n,\alpha }^{^{\Gamma ~\ast }}$ is
approximated as accurately as desired by $\left( \Gamma \left( \hat{R}%
_{n}^{p,\ast }\right) \right) _{B(1-\alpha )}$, the $B\left( 1-\alpha
\right)-$th order statistic from $B$ replicates $\left\{ \left(\Gamma \left( 
\hat{R}_{n}^{p,\ast }\right)\right)_l \right\} _{l=1}^{B}$ of $\Gamma \left( 
\hat{R}_{n}^{p,\ast }\right) $.

The next theorem establishes the asymptotic validity of the multiplier
bootstrap procedure proposed above.

\begin{theorem}
\label{bootstrap} Assume Assumptions \ref{ass1}-\ref{ass3}. Then, 
\begin{equation*}
\hat{R}_{n}^{p\ast }\underset{\ast }{\Rightarrow }R_{\infty }^{p}\quad \text{%
a.s.},
\end{equation*}%
where $R_{\infty }^{p}$ is the Gaussian process defined in Theorem 1, and $%
\underset{\ast }{\Rightarrow }$ denotes the weak convergence under the
bootstrap law, i.e. conditional on the original sample $\{(D_{i},X_{i}^{%
\prime })^{\prime }\}_{i=1}^{n}$. Additionally, for any continuous
functional $\Gamma (\cdot )$ from $l^\infty(\Pi)$ to $\mathbb{R}$, we have $%
\Gamma \left( \hat{R}_{n}^{p,\ast }\right) \underset{\ast }{\overset{d}{%
\rightarrow }}\Gamma \left( R_{\infty }^{p}\right) $ a.s. under the
bootstrap law.
\end{theorem}

\section{Monte Carlo simulation study\label{MC}}

In this section, we conduct a series of Monte Carlo experiments in order to
study the finite sample properties of our proposed projection-based tests.
In particular, we compare our Cram\'{e}r-von Mises and Kolmogorov-Smirnov
tests $CvM_{n}$ and $KS_{n}$ given in (\ref{cvm}) and (\ref{ks}) to $(i)$
the \cite{Shaikh2009}'s test,%
\begin{equation*}
T_{n}\left( h_{n}\right) =\sqrt{\frac{n-1}{n}}\frac{nh_{n}^{1/2}\hat{V}%
_{n}\left( h_{n}\right) }{\sqrt{\hat{\Sigma}_{n}\left( h_{n}\right) }},
\end{equation*}%
where $\hat{V}_{n}\left( h_{n}\right) $ is given in (\ref{shaikh}) and 
\begin{equation*}
\hat{\Sigma}_{n}\left( h_{n}\right) =\frac{2}{n\left( n-1\right) }%
\sum_{i=1}^{n}\sum_{j=1,j\not=i}^{n}\frac{1}{h_{n}}K^{2}\left( \frac{q\left(
X_{i},\hat{\theta}_{n}\right) -q\left( X_{j},\hat{\theta}_{n}\right) }{h_{n}}%
\right) \varepsilon _{i}^{2}\left( \hat{\theta}_{n}\right) \varepsilon
_{j}^{2}\left( \hat{\theta}_{n}\right) ;
\end{equation*}%
the analogues of $CvM_{n}$ and $KS_{n}$ based on either $\left( ii\right) $
the unprojected process $\hat{R}_{n}(u)$ given in (\ref{R1})\footnote{%
For conciseness, we do not discuss the asymptotic properties of tests based
on the unprojected empirical process $\hat{R}_{n}(u)$ in the main text.
However, most of the asymptotic properties follow from arguments analogous
to those we used to study $\hat{R}_{n}^{p}(u)$.}, or on $\left( iii\right) $
the traditional empirical process%
\begin{equation}
\hat{R}_{n}^{trad}\left( x\right) =\frac{1}{\sqrt{n}}\sum_{i=1}^{n}%
\varepsilon _{i}(\hat{\theta}_{n})1\left\{ X_{i}\leq x\right\} .
\label{rtrad}
\end{equation}%
We also compare our proposal with $\left( iv\right) $ balancing tests based
on the normalized IPW estimators 
\begin{equation}
\widehat{ATE}_{n}\left( X^{j}\right) =\frac{1}{n}\sum_{i=1}^{n}\left( \frac{%
w_{1,i}}{\bar{w}_{1,n}}-\frac{w_{0,i}}{\bar{w}_{0,n}}\right) X_{i}^{j},
\label{ate.x}
\end{equation}%
where $w_{1,i}=D_{i}/q\left( X_{i},\hat{\theta}_{n}\right) $, $%
w_{0,i}=\left( 1-D_{i}\right) /\left( 1-q\left( X_{i},\hat{\theta}%
_{n}\right) \right) $, $\bar{w}_{d,n}$ is the sample mean of $w_{d,i},$ $%
d=\left\{ 0,1\right\} $, and $X^{j}$ is the $j$-th element of $d$%
-dimensional vector $X=\left( X_{1},\dots ,X_{d}\right) ^{\prime }$. We
consider two different test statistics for the balancing tests: the Wald
test, and the maximum of the $d$ marginal two-sided $t$-tests.

Critical values (CVs) for the $CvM_{n}$ and $KS_{n}$ tests are obtained
using the multiplier bootstrap procedure described in Section \ref{boots},
whereas for the ICM tests in $\left( ii\right) $ and $\left( iii\right) $ we
use the bootstrap procedure described in \cite{Stute1998b}. For $T_{n}\left(
h_{n}\right)$ test, we use one-sided CVs from the standard normal
distribution. CVs for the Wald test are from the chi-squared distribution
with $d$ degrees of freedom. For the $t$-test, we consider Bonferroni
corrected CVs based on the standard normal distribution. Note that the
Bonferroni correction is necessary to address the multiple testing problem.

We consider sample sizes $n$ equal to $100$, $200$, $400$, $600$, $800$ and $%
1,000$. For each design, we consider $1,000$ Monte Carlo experiments. The $%
\left\{ V_{i}\right\} _{i=}^{n}$ used in the bootstrap implementations are
independently generated as $V$ with $\mathbb{P}\left( V=1-\kappa \right)
=\kappa /\sqrt{5}$ and $\mathbb{P}\left( V=\kappa \right) =1-\kappa /\sqrt{5}
$, where $\kappa =\left( \sqrt{5}+1\right) /2$, as proposed by \cite%
{Mammen1993}. The bootstrapped critical values are approximated using $B=999$
bootstrap replications. To compute \cite{Shaikh2009}'s test, we use the
standard normal kernel%
\begin{equation*}
K\left( u\right) =\frac{1}{\sqrt{2\pi }}\exp \left( -\frac{u^{2}}{2}\right) 
\text{.}
\end{equation*}%
Following \cite{Shaikh2009}, the bandwidth sequence $h_{n}$ is chosen to be
equal to $cn^{-1/8}$ for $c$ equal to 0.01, 0.05, 0.10 and 0.15. We choose
different $c$'s to assess how sensitive \cite{Shaikh2009}'s test may be with
respect to the bandwidth $h_{n}$.

\subsection{Simulation 1\label{simulation1}}

We first consider the following data generating processes (DGPs):%
\begin{align*}
DGP1.~D^{\ast }& =\frac{\left( X_{1}+X_{2}\right) }{3}-\varepsilon ; \\[1ex]
DGP2.~D^{\ast }& =-1+\frac{\left( X_{1}+X_{2}+X_{1}X_{2}\right) }{3}%
-\varepsilon ; \\[1ex]
DGP3.~D^{\ast }& =-0.2+\frac{\left( X_{1}^{2}-X_{2}^{2}\right) }{2}%
-\varepsilon ; \\[1ex]
DGP4.~D^{\ast }& =\frac{\left( 0.1+X_{1}/3\right) }{\exp \left( \left.
\left( X_{1}+X_{2}\right) \right/ 3\right) }-\varepsilon ; \\[1ex]
DGP5.~D^{\ast }& =\frac{\left( -0.8+\left( X_{1}+X_{2}+X_{1}X_{2}\right)
/3\right) }{\exp \left( \left. 0.2+\left( X_{1}+X_{2}\right) \right/
3\right) }-\varepsilon .
\end{align*}%
For each of these five DGPs, $D=1\left\{ D^{\ast }>0\right\} ,\varepsilon 
%TCIMACRO{\TeXButton{indep}{\independent}}%
%BeginExpansion
\independent%
%EndExpansion
\left( X_{1},X_{2}\right) $, where $X_{1}=Z_{1}$, $X_{2}=\left(
Z_{1}+Z_{2}\right) /\sqrt{2}$, and $Z_{1}$, $Z_{2},$ and $\varepsilon $ are
independent standard normal random variables. All the DGPs considered have
the propensity score bounded away from zero and one. Finally, for each of
these DGPs we consider the potential outcomes%
\begin{equation*}
Y\left( 1\right) =2m_{1}(X)+u\left( 1\right) \quad \text{and}\quad Y\left(
0\right) =m_{1}\left( X\right) +u\left( 0\right) ,\ 
\end{equation*}%
where $m_{1}\left( X\right) =1+X_{1}+X_{2}$, $u\left( 1\right) $ and $%
u\left( 0\right) $ are independent normal random variables with mean zero
and variance $0.1$. The observed outcome is $Y=DY\left( 1\right) +\left(
1-D\right) Y\left( 0\right) $, and the true $ATE$ is 1. Although these
outcome equations are not necessary to assess the size and power properties
of the tests, they can be used to assess the utility of our proposed tests
to distinguish between \textquotedblleft good\textquotedblright\ and
\textquotedblleft bad\textquotedblright\ estimates of the $ATE$.

Let $X=\left( 1,X_{1},X_{2}\right) ^{\prime }$. For $DGP1-DGP5$, the $H_{0}$
considered is 
\begin{equation*}
H_{0}:\exists \theta _{0}=\left( \beta _{0},\beta _{1},\beta _{2}\right)
^{\prime }\in \Theta: \mathbb{E}\left[ D|\Phi \left( X^{\prime }\theta
_{0}\right) \right] =\Phi \left( X^{\prime }\theta _{0}\right) ~a.s.,
\end{equation*}%
where $\Phi (\cdot )$ is the cumulative distribution function (CDF) of the
standard normal distribution. We estimate $\theta _{0}$ using the Probit ML,
i.e. 
\begin{equation*}
\hat{\theta}_{n}=\arg \max_{\theta\in\Theta }\sum_{i=1}^{n}D_{i}\ln \left(
\Phi \left( X_{i}^{\prime }\theta \right) \right) +\left( 1-D_{i}\right) \ln
\left( 1-\Phi \left( X_{i}^{\prime }\theta \right) \right).
\end{equation*}%
Clearly, $DGP1$ falls under $H_{0}$, whereas $DGP2-DGP5$ fall under $H_{1}$.
Note that $D$ follows a heteroskedastic Probit model in $DGP4$ and $DGP5$.

The simulation results are presented in Table \ref{tab:sim1}. We report
empirical rejection frequencies at the $5\%$ significance level. Results for
10\% and 1\% significance levels are similar and are available upon request.
We also report the bias of the normalized IPW estimator 
\begin{equation}
\widehat{ATE_{n}}=\frac{1}{n}\sum_{i=1}^{n}\left( \frac{w_{1,i}}{\bar{w}%
_{1,n}}-\frac{w_{0,i}}{\bar{w}_{0,n}}\right) Y_{i},  \label{ate}
\end{equation}%
and the average length and coverage of its estimated 95\% confidence
interval based on its asymptotically normal approximation (assuming that the
propensity score model is correctly specified).

We first analyze the size of our test. From the results of $DGP1$, we find
that the actual finite sample size of both $KS_{n}$ and $CvM_{n}$ tests is
close to their nominal size, even when the sample size is as small as 100.
The same holds for the other ICM-type tests. On the other hand, we find that 
\cite{Shaikh2009}'s test is, in general, conservative, and sensitive to the
choice of bandwidth. For instance, when $c=0.15$, the empirical size is
close to zero even with $n=1,000$. On the other hand, with $c=0.01$, the
empirical size of \cite{Shaikh2009}'s test is closer to the nominal value.
In terms of traditional balancing tests, we find that Wald test and
Bonferroni-corrected $t$-test do not control size. Such a drawback is due to
the random denominator being relatively close to zero: when we trim
observations with estimated propensity score outside the $[0.05,0.95]$
range, we find that classical balancing tests can control size. We report
these results in Table \ref{tab:mc1trim} in the Appendix B. Finally, note
that when the propensity score is correctly specified, the bias of the $%
\widehat{ATE_n}$ estimator in (\ref{ate}) is small, the length of the 95\%
confidence interval reduces as sample size increases, but the coverage
probability is smaller than its nominal value even when $n=1,000$. However,
as we show in Table \ref{tab:mc1trim} in the Appendix B, such undercoverage
disappears when we trim observations with extreme estimated propensity
scores.

\afterpage{
\begin{landscape}
\begin{table}[htbp]
\caption{Monte Carlo results under designs $DGP1$-$DGP5$}
\label{tab:sim1}\centering
\centering\begin{adjustbox}{ max width=1\linewidth, max totalheight=1\textheight, keepaspectratio}
\begin{threeparttable}
    \begin{tabular}{ccccccccccccccccc} \hline
    \toprule
    DGP   & $n$     & \multicolumn{1}{c}{$CvM_{n}$} & \multicolumn{1}{c}{$KS_{n}$} & \multicolumn{1}{c}{$T_{n}(0.01)$} & \multicolumn{1}{c}{$T_{n}(0.05)$} & \multicolumn{1}{c}{$T_{n}(0.10)$} & \multicolumn{1}{c}{$T_{n}(0.15)$} & \multicolumn{1}{l}{Max-$t$} & \multicolumn{1}{l}{$Wald$} & \multicolumn{1}{c}{$CvM_{n}^{unp}$} & \multicolumn{1}{c}{$KS_{n}^{unp}$} & \multicolumn{1}{c}{$CvM_{n}^{trad}$} & \multicolumn{1}{c}{$KS_{n}^{trad}$} & \multicolumn{1}{c}{Bias} & \multicolumn{1}{c}{CI length} & \multicolumn{1}{c}{Coverage} \\ \hline
    \midrule
    1     & 100   & 5.00  & 5.60  & 4.90  & 2.30  & 0.80  & 0.20  & 8.90  & 6.90  & 5.40  & 5.00  & 5.40  & 6.10  & 0.11  & 1.36  & 85.60 \\
    1     & 200   & 4.40  & 4.60  & 4.90  & 2.30  & 0.60  & 0.20  & 8.80  & 8.70  & 4.00  & 4.50  & 4.40  & 3.90  & 0.02  & 1.04  & 90.00 \\
    1     & 400   & 5.10  & 5.80  & 4.10  & 2.40  & 1.10  & 0.40  & 12.10 & 12.40 & 4.70  & 5.30  & 4.70  & 4.80  & 0.01  & 0.76  & 90.80 \\
    1     & 600   & 5.20  & 5.70  & 4.30  & 1.80  & 0.70  & 0.40  & 13.20 & 14.50 & 5.00  & 5.30  & 4.60  & 5.50  & 0.01  & 0.64  & 87.20 \\
    1     & 800   & 5.30  & 4.30  & 4.50  & 1.50  & 0.50  & 0.40  & 13.80 & 16.30 & 4.90  & 4.50  & 6.20  & 4.70  & 0.01  & 0.56  & 89.50 \\
    1     & 1000  & 5.00  & 5.90  & 4.50  & 2.90  & 1.20  & 0.80  & 13.50 & 15.10 & 5.10  & 4.90  & 5.70  & 5.00  & 0.01  & 0.52  & 90.70 \\ \hline
    2     & 100   & 28.90 & 26.40 & 6.70  & 7.60  & 7.70  & 6.20  & 14.00 & 26.80 & 17.90 & 18.70 & 16.10 & 16.80 & -1.76 & 6.48  & 89.00 \\
    2     & 200   & 61.50 & 55.30 & 7.30  & 17.30 & 20.80 & 18.70 & 13.80 & 27.00 & 41.60 & 42.60 & 35.50 & 30.50 & -2.75 & 6.39  & 78.70 \\
    2     & 400   & 90.60 & 84.30 & 18.50 & 44.60 & 50.40 & 50.40 & 41.60 & 45.00 & 79.80 & 74.80 & 71.50 & 60.60 & -3.50 & 5.97  & 29.30 \\
    2     & 600   & 98.90 & 96.30 & 33.60 & 66.60 & 74.00 & 75.80 & 75.50 & 75.50 & 94.90 & 93.40 & 89.80 & 82.80 & -3.95 & 5.84  & 9.80 \\
    2     & 800   & 99.70 & 99.10 & 50.40 & 83.50 & 89.20 & 89.70 & 88.40 & 89.50 & 99.10 & 98.00 & 97.70 & 93.80 & -4.14 & 5.78  & 3.60 \\
    2     & 1000  & 100.00 & 100.00 & 67.40 & 93.90 & 96.10 & 96.80 & 95.80 & 96.00 & 99.90 & 99.90 & 99.40 & 98.40 & -4.30 & 5.44  & 0.80 \\ \hline
    3     & 100   & 32.80 & 26.10 & 6.20  & 5.30  & 1.50  & 0.40  & 0.10  & 0.10  & 33.60 & 25.00 & 15.90 & 18.10 & 0.01  & 0.87  & 95.90 \\
    3     & 200   & 59.10 & 49.90 & 18.70 & 16.50 & 6.40  & 1.10  & 0.00  & 0.00  & 59.00 & 49.60 & 40.80 & 38.80 & -0.01 & 0.57  & 94.70 \\
    3     & 400   & 69.20 & 64.10 & 38.40 & 28.60 & 9.20  & 2.40  & 0.00  & 0.00  & 69.70 & 63.70 & 83.00 & 80.30 & 0.00  & 0.39  & 94.70 \\
    3     & 600   & 79.70 & 74.60 & 56.60 & 40.90 & 14.70 & 3.80  & 0.00  & 0.00  & 80.10 & 75.30 & 98.70 & 97.10 & 0.00  & 0.32  & 95.70 \\
    3     & 800   & 81.00 & 77.70 & 63.40 & 42.30 & 15.20 & 4.10  & 0.00  & 0.00  & 81.30 & 77.80 & 99.70 & 99.70 & 0.00  & 0.27  & 96.20 \\
    3     & 1000  & 83.30 & 81.30 & 71.00 & 48.90 & 17.70 & 3.90  & 0.10  & 0.00  & 82.90 & 81.00 & 100.00 & 100.00 & 0.00  & 0.24  & 95.80 \\ \hline
    4     & 100   & 15.20 & 12.20 & 4.40  & 1.50  & 1.10  & 0.50  & 0.30  & 0.30  & 15.60 & 12.00 & 20.00 & 11.40 & 0.17  & 0.90  & 90.70 \\
    4     & 200   & 35.00 & 27.00 & 5.10  & 7.30  & 6.50  & 3.50  & 1.10  & 0.60  & 37.50 & 26.80 & 42.00 & 25.10 & 0.16  & 0.60  & 80.60 \\
    4     & 400   & 64.70 & 53.80 & 8.10  & 17.20 & 21.10 & 17.00 & 4.10  & 2.20  & 70.30 & 51.20 & 74.60 & 46.30 & 0.16  & 0.42  & 69.20 \\
    4     & 600   & 85.30 & 75.40 & 13.40 & 33.00 & 39.00 & 35.70 & 8.10  & 5.20  & 88.30 & 72.50 & 90.50 & 69.90 & 0.16  & 0.34  & 55.50 \\
    4     & 800   & 92.50 & 86.70 & 19.00 & 50.00 & 60.40 & 59.40 & 9.50  & 7.40  & 94.80 & 85.00 & 96.70 & 83.70 & 0.16  & 0.29  & 41.90 \\
    4     & 1000  & 96.50 & 93.80 & 29.80 & 63.20 & 74.00 & 75.40 & 12.80 & 10.00 & 97.60 & 92.00 & 99.20 & 90.40 & 0.16  & 0.26  & 32.50 \\ \hline
    5     & 100   & 11.00 & 7.80  & 5.00  & 2.20  & 1.30  & 0.60  & 22.40 & 22.40 & 9.20  & 6.90  & 9.20  & 5.70  & 0.04  & 2.70  & 68.30 \\
    5     & 200   & 16.40 & 13.70 & 4.90  & 2.70  & 2.30  & 1.60  & 29.10 & 32.40 & 13.80 & 9.70  & 13.00 & 7.80  & -0.31 & 2.74  & 68.00 \\
    5     & 400   & 26.70 & 23.00 & 5.90  & 5.70  & 4.70  & 3.90  & 24.10 & 29.60 & 23.50 & 15.80 & 22.70 & 11.10 & -0.72 & 3.04  & 72.60 \\
    5     & 600   & 35.80 & 32.80 & 5.30  & 8.70  & 9.90  & 8.40  & 18.50 & 24.50 & 35.20 & 21.70 & 32.40 & 15.50 & -0.93 & 3.26  & 77.40 \\
    5     & 800   & 45.00 & 39.30 & 5.70  & 11.80 & 14.30 & 12.90 & 17.50 & 21.80 & 44.00 & 28.60 & 38.80 & 19.10 & -1.02 & 3.23  & 77.80 \\
    5     & 1000  & 56.00 & 49.20 & 7.80  & 17.00 & 20.70 & 19.30 & 16.60 & 21.70 & 57.20 & 38.30 & 52.40 & 27.00 & -1.06 & 3.21  & 76.30 \\ \hline 
        \bottomrule
    \end{tabular}    \begin{tablenotes}[para,flushleft]
\small{
Note: Simulations based on 1,000 Monte Carlo experiments. ``$CvM_{n}$'' and ``$KS_{n}$'' stand for our proposed Cram\'{e}r-von Mises and Kolmogorov-Smirnov tests. ``$T_{n}(c)$'' stands for \cite{Shaikh2009}'s test, with bandwidth $h_{n}=cn^{-1/8}$. 
``Max-$t$'' and ``$Wald$'' stand for, respectively, the Bonferroni-corrected Max-$t$-test, and Wald balancing test based on $\widehat{ATE}_{n}\left( X^{j}\right)$ defined in (\ref{ate.x}).  ``$CvM_{n}^{unp}$'' and ``$KS_{n}^{unp}$'' are the  Cram\'{e}r-von Mises and Kolmogorov-Smirnov tests based on the unprojected empirical process $\hat{R}_{n}(u)$ defined in (\ref{R1}), whereas ``$CvM_{n}^{trad}$'' and ``$KS_{n}^{trad}$'' are defined analogously, but based on $\hat{R}_{n}^{trad}\left( x\right)$ as defined in (\ref{rtrad}). Finally, ``Bias'', ``CI length', and ``Coverage'' stands for the average simulated bias, estimated 95\% confidence interval length, and 95\% coverage probability for the $ATE$ estimator $\widehat{ATE_{n}}$ as defined in (\ref{ate}). All entries are proportions of rejections at 5\% level, in percentage points, except ``Bias'' ,``CI length'', and ``Coverage'' (measure in percentage points), which are as described above. See the main text for further details.}
\end{tablenotes}
\end{threeparttable}
\end{adjustbox}
\end{table}
\end{landscape}
}

Note that when the propensity score is misspecified in $DGP2-DGP5$, the $ATE$
estimator (\ref{ate}) can be severely biased, and its confidence interval
can be \textquotedblleft too small\textquotedblright\ when the propensity
score is misspecified, leading to severe undercoverage\footnote{%
In $DGP3$, the bias, confidence interval length, and coverage of the $ATE$
estimator are good. However, it is important to have in mind that we
consider only one particular DGP for the $ATE$, and that such
\textquotedblleft robust\textquotedblright\ results may not translate to
other DGPs.}. Thus, detecting propensity score misspecifications can prevent
misleading inference about $ATE$. Our proposed $KS_{n}$ and $CvM_{n}$ tests
perform admirably well in such a task, particularly in moderate sample
sizes. In these scenarios, $CvM_{n}$ performs slightly better than $KS_{n}$.
Looking at the results from \cite{Shaikh2009}'s test, we note that bandwidth
choices can play an important role, and the choice of the \textquotedblleft
best\textquotedblright\ bandwidth $h_n$ via $c$ varies across DGPs. Perhaps,
what is more important to emphasize in terms of power is that in all
alternative hypotheses and sample sizes analyzed, our projection based tests
have higher power than \cite{Shaikh2009}'s test, regardless of the bandwidth
choice. Our proposed tests also dominate the balancing tests, as balancing
tests have little to no power in all DGPs considered but $DGP2$, even with $%
n=1,000$. Finally, the results in Table \ref{tab:sim1} show that
projection-based tests perform either better or as well as the other ICM
tests in the DGPs considered. Among the considered DGPs, $DGP3$ is the only
one where some existing specification testing has higher power than our
proposed procedure. For this particular DGP, ICM type tests $KS^{trad}_n$
and $CvM^{trad}_n$ based on (\ref{rtrad}) have higher power than our
proposed tests when sample size $n$ is large. Given the discussion in
Section \ref{powersec} and the fact that none of the ICM tests are strictly
better than the others uniformly over the space of alternatives, such a
finding does not come with a surprise.

\subsection{Simulation 2\label{simulation2}}

In this simulation, we push forward the dimensionality of the covariates to
see how our proposed tests and the other alternative tests perform in
scenarios with 10 continuous covariates. To investigate further this issue,
we consider the following five DGPs:%
\begin{align*}
DGP6.~D^{\ast }& =-\frac{\sum_{j=1}^{10}X_{j}}{6}-\varepsilon ,; \\[1ex]
DGP7.~D^{\ast }& =-1-\frac{\sum_{j=1}^{10}X_{j}}{10}+\frac{X_{1}X_{2}}{2}%
-\varepsilon ,; \\[1ex]
DGP8.~D^{\ast }& =-1-\frac{\sum_{j=1}^{10}X_{j}}{10}+\frac{%
X_{1}\sum_{k=2}^{5}X_{k}}{4}-\varepsilon ,; \\[1ex]
DGP9.~D^{\ast }& =-1.5-\frac{\sum_{j=1}^{10}X_{j}}{6}+\frac{%
\sum_{k=1}^{10}X_{k}^{2}}{10}-\varepsilon ; \\[1ex]
DGP10.~D^{\ast }& =\frac{-0.1+0.1\sum_{j=1}^{5}X_{j}}{\exp \left(
-0.2\sum_{k=1}^{10}X_{j}\right) }-\varepsilon ,,
\end{align*}%
where $X_{1}$, $X_{2}$ and $\varepsilon $ are defined as before, $\left\{
X_{i}\right\} _{k=3}^{10}$ are independent standard normal random variables, 
$D=1\left\{ D^{\ast }>0\right\} ,$ and $\varepsilon 
%TCIMACRO{\TeXButton{indep}{\independent}}%
%BeginExpansion
\independent%
%EndExpansion
X$, with $X=\left( 1,X_{1},X_{2},\dots ,X_{10}\right) ^{\prime }$. For each
of these DGPs we consider the potential outcomes%
\begin{equation*}
Y\left( 1\right) =2m_{2}(X)+u\left( 1\right) \quad \text{and}\quad Y\left(
0\right) =m_{2}\left( X\right) +u\left( 0\right) ,\ 
\end{equation*}%
where $m_{2}\left( X\right) =1+\sum_{j=1}^{10}X_{j}$, $u\left( 1\right) $
and $u\left( 0\right) $ are independent normal random variables with mean
zero and variance $0.1$. The observed outcome is $Y=DY\left( 1\right)
+\left( 1-D\right) Y\left( 0\right) $, and the true $ATE$ is 1.

For $DGP6-DGP10$, the $H_{0}$ considered is 
\begin{equation}
H_{0}:\exists \theta
_{0}=(\beta_0,\beta_1,\beta_2,\ldots,\beta_{10})^{\prime }\in \Theta:\mathbb{%
E}\left[ D|\Phi \left( X^{\prime }\theta _{0}\right) \right] =\Phi \left(
X^{\prime }\theta _{0}\right) ~a.s..  \label{nullk}
\end{equation}%
We estimate $\theta _{0}$ by ML$.$ Note that $DGP6$ falls under $H_{0},$
whereas $DGP7$-$DGP10$ fall under $H_{1}$. The simulation results for $DGP6$-%
$DGP10$ are presented in Table \ref{tab:sim2}.

As before, we first discuss the size properties of the tests. From the
results of $DGP6$, we find that $KS_{n}$ and $CvM_{n}$ tests are oversized
when $n$ $=100$, but as sample size $n$ increases, the empirical size gets
closer to its nominal value. The same holds true for ICM tests based on the
unprojected process (\ref{R1}). \cite{Shaikh2009}'s test tends to be
conservative (with the exception when $c=0.01),$ and sensitive to the
bandwidth choice. The traditional balancing tests, and the ICM tests based
on (\ref{rtrad}) are conservative, reflecting the \textquotedblleft curse of
dimensionality\textquotedblright . Finally, note that when the propensity
score is correctly specified, the finite sample properties of the $ATE$
estimator (\ref{ate}) are good: the bias and the length of the 95\%
confidence interval get smaller when sample size increases, and the coverage
probability is relatively close to its nominal value. When one trims
observations with extreme estimated propensity score, these properties
further improve; see Table \ref{tab:mc2trim} in the Appendix B.

Note that when the propensity score is misspecified, the $ATE$ estimator (%
\ref{ate}) can be severely biased, and inference can be unreliable. Thus,
tests with higher power to detect such misspecifications can prevent one to
make misleading conclusions about the effectiveness of a given policy. What
is clear from Table \ref{tab:sim2} is that, regardless of the sample size
and bandwidth considered, \cite{Shaikh2009}'s test seems to have little to
no power to detect the alternatives described in $DGP7$, $DGP9$, and $DGP10$%
. For $DGP8$, the maximum power for their test is approximately 55\% when $%
n=1,000$ and $c=0.1.$ However, with $c=0.01,$ the power of \cite{Shaikh2009}%
's test reduces to approximately 30\%, highlighting again how important (and
non-trivial) is to \textquotedblleft appropriately\textquotedblright\ choose
the bandwidth. In sharp contrast with \cite{Shaikh2009}'s test, note that
for moderately sized samples, our proposed $KS_{n}$ and $CvM_{n}$ tests have
non-trivial power to detect all the alternatives. Our projection-based tests
seem to dominate the other tests in the scenarios considered. Note that the
traditional balancing tests have no power to detect the alternatives
considered. Finally, ICM tests based on the traditional empirical process (%
\ref{rtrad}) have substantially less power than our proposed tests,
reflecting the cost of the \textquotedblleft curse of
dimensionality\textquotedblright . The power gains from using the
projection-based process (\ref{emp-process}) instead of the unprojected
process (\ref{R1}) can also be noted.

\afterpage{
\begin{landscape}
\begin{table}[htbp]
\caption{Monte Carlo results under designs $DGP6$-$DGP10$}
\label{tab:sim2}\centering
\centering\begin{adjustbox}{ max width=1\linewidth, max totalheight=1\textheight, keepaspectratio}
\begin{threeparttable}
    \begin{tabular}{ccccccccccccccccc} \hline
    \toprule
    DGP   & $n$     & \multicolumn{1}{c}{$CvM_{n}$} & \multicolumn{1}{c}{$KS_{n}$} & \multicolumn{1}{c}{$T_{n}(0.01)$} & \multicolumn{1}{c}{$T_{n}(0.05)$} & \multicolumn{1}{c}{$T_{n}(0.10)$} & \multicolumn{1}{c}{$T_{n}(0.15)$} & \multicolumn{1}{l}{Max-$t$} & \multicolumn{1}{l}{$Wald$} & \multicolumn{1}{c}{$CvM_{n}^{unp}$} & \multicolumn{1}{c}{$KS_{n}^{unp}$} & \multicolumn{1}{c}{$CvM_{n}^{trad}$} & \multicolumn{1}{c}{$KS_{n}^{trad}$} & \multicolumn{1}{c}{Bias} & \multicolumn{1}{c}{CI length} & \multicolumn{1}{c}{Coverage} \\ \hline
    \midrule
     
    6     & 100   & 8.50  & 9.40  & 5.80  & 2.50  & 1.00  & 0.20  & 1.10  & 1.60  & 7.80  & 9.40  & 1.10  & 1.70  & -0.29 & 3.71  & 89.40 \\
    6     & 200   & 5.70  & 6.10  & 4.00  & 2.50  & 1.00  & 0.40  & 0.90  & 0.30  & 5.70  & 6.50  & 2.30  & 3.90  & -0.08 & 2.10  & 91.00 \\
    6     & 400   & 5.60  & 6.00  & 5.10  & 2.40  & 0.90  & 0.30  & 2.80  & 0.70  & 5.80  & 6.30  & 3.30  & 4.50  & -0.02 & 1.41  & 91.90 \\
    6     & 600   & 5.00  & 5.20  & 4.90  & 1.70  & 0.60  & 0.20  & 3.40  & 1.60  & 3.70  & 5.00  & 3.80  & 3.70  & -0.02 & 1.11  & 90.00 \\
    6     & 800   & 5.40  & 5.20  & 5.10  & 2.70  & 1.40  & 0.70  & 5.10  & 3.40  & 4.60  & 5.00  & 2.60  & 4.20  & -0.01 & 0.97  & 93.30 \\
    6     & 1000  & 4.90  & 6.00  & 4.50  & 2.80  & 1.10  & 0.50  & 4.90  & 5.20  & 5.70  & 5.80  & 3.70  & 5.40  & -0.01 & 0.84  & 92.30 \\ \hline
    7     & 100   & 9.50  & 10.60 & 4.10  & 1.60  & 0.70  & 0.00  & 0.60  & 5.30  & 6.80  & 8.40  & 2.50  & 3.80  & 0.22  & 6.97  & 96.60 \\
    7     & 200   & 15.40 & 14.90 & 5.10  & 3.30  & 2.00  & 1.10  & 0.70  & 1.20  & 13.60 & 13.90 & 2.70  & 6.00  & 0.59  & 4.43  & 99.10 \\
    7     & 400   & 26.40 & 24.20 & 5.40  & 6.60  & 6.80  & 5.20  & 0.60  & 0.30  & 23.00 & 19.90 & 4.60  & 7.90  & 0.61  & 2.66  & 99.10 \\
    7     & 600   & 37.20 & 33.00 & 9.10  & 10.70 & 11.10 & 8.70  & 0.10  & 0.00  & 33.70 & 30.10 & 8.30  & 9.30  & 0.56  & 1.97  & 96.60 \\
    7     & 800   & 49.50 & 46.50 & 9.90  & 16.70 & 17.00 & 13.90 & 0.10  & 0.00  & 46.40 & 37.70 & 13.80 & 13.70 & 0.59  & 1.69  & 89.30 \\
    7     & 1000  & 57.20 & 52.90 & 11.90 & 22.90 & 24.30 & 21.60 & 0.20  & 0.20  & 55.60 & 47.50 & 17.70 & 16.20 & 0.60  & 1.51  & 78.10 \\ \hline
    8     & 100   & 9.50  & 10.60 & 5.70  & 1.30  & 0.80  & 0.20  & 2.00  & 14.30 & 8.20  & 9.90  & 2.20  & 3.70  & 0.61  & 10.68 & 96.20 \\
    8     & 200   & 18.30 & 18.70 & 4.40  & 4.20  & 3.60  & 2.80  & 1.00  & 4.40  & 14.80 & 15.50 & 3.40  & 5.30  & 1.12  & 6.58  & 99.40 \\
    8     & 400   & 46.50 & 41.80 & 8.10  & 16.00 & 15.80 & 11.60 & 0.50  & 0.90  & 41.70 & 37.10 & 9.10  & 9.10  & 1.36  & 4.44  & 98.50 \\
    8     & 600   & 66.30 & 62.10 & 15.80 & 27.50 & 28.10 & 24.20 & 0.20  & 0.50  & 60.60 & 54.20 & 20.80 & 16.10 & 1.33  & 3.42  & 88.30 \\
    8     & 800   & 79.90 & 74.40 & 23.30 & 42.70 & 45.60 & 40.00 & 0.50  & 0.60  & 74.90 & 68.30 & 30.90 & 22.30 & 1.34  & 2.91  & 66.00 \\
    8     & 1000  & 87.00 & 82.10 & 31.30 & 52.70 & 55.00 & 51.20 & 2.10  & 0.30  & 82.40 & 77.70 & 42.20 & 25.40 & 1.30  & 2.51  & 44.00 \\ \hline
    9     & 100   & 10.00 & 12.00 & 3.50  & 2.40  & 0.40  & 0.20  & 1.70  & 6.80  & 10.10 & 12.40 & 1.20  & 2.50  & 0.06  & 7.76  & 88.30 \\
    9     & 200   & 15.10 & 15.10 & 4.70  & 3.70  & 1.80  & 0.80  & 1.60  & 4.10  & 12.40 & 12.00 & 4.00  & 7.10  & 0.60  & 5.02  & 94.60 \\
    9     & 400   & 30.40 & 26.80 & 4.50  & 3.10  & 2.80  & 2.90  & 0.80  & 2.70  & 21.30 & 17.00 & 8.50  & 12.00 & 0.97  & 3.70  & 97.30 \\
    9     & 600   & 44.10 & 36.10 & 5.70  & 6.10  & 6.70  & 6.00  & 0.50  & 3.30  & 32.50 & 25.30 & 17.10 & 20.10 & 1.00  & 3.01  & 89.20 \\
    9     & 800   & 57.60 & 50.20 & 6.60  & 8.80  & 11.00 & 10.40 & 0.70  & 3.00  & 46.80 & 35.60 & 25.40 & 29.30 & 1.03  & 2.63  & 78.60 \\
    9     & 1000  & 70.50 & 57.50 & 5.30  & 11.30 & 15.70 & 16.60 & 0.80  & 2.40  & 59.30 & 45.60 & 35.60 & 39.80 & 1.07  & 2.45  & 67.90 \\ \hline
    10    & 100   & 7.50  & 8.30  & 4.10  & 1.30  & 0.60  & 0.30  & 0.00  & 0.00  & 6.80  & 8.70  & 4.00  & 3.90  & -0.09 & 2.37  & 97.30 \\
    10    & 200   & 10.00 & 10.10 & 5.10  & 1.80  & 0.80  & 0.30  & 0.00  & 0.00  & 10.30 & 9.30  & 4.40  & 4.90  & -0.10 & 1.26  & 96.50 \\
    10    & 400   & 19.20 & 16.70 & 5.20  & 3.80  & 2.80  & 1.80  & 0.00  & 0.00  & 20.60 & 14.90 & 6.90  & 8.60  & -0.10 & 0.81  & 93.60 \\
    10    & 600   & 36.20 & 28.10 & 4.60  & 7.30  & 6.80  & 4.80  & 0.10  & 0.00  & 37.70 & 27.70 & 10.10 & 8.20  & -0.11 & 0.65  & 91.40 \\
    10    & 800   & 50.20 & 38.90 & 7.70  & 11.30 & 13.20 & 11.10 & 0.10  & 0.00  & 52.80 & 38.80 & 15.20 & 12.60 & -0.12 & 0.55  & 87.50 \\
    10    & 1000  & 64.30 & 51.70 & 8.20  & 18.20 & 22.30 & 18.90 & 0.10  & 0.00  & 67.00 & 51.40 & 21.00 & 13.90 & -0.12 & 0.49  & 83.10 \\ \hline
        \bottomrule
    \end{tabular}    \begin{tablenotes}[para,flushleft]
\small{
Note: Simulations based on 1,000 Monte Carlo experiments. ``$CvM_{n}$'' and ``$KS_{n}$'' stand for our proposed Cram\'{e}r-von Mises and Kolmogorov-Smirnov tests. ``$T_{n}(c)$'' stands for \cite{Shaikh2009}'s test, with bandwidth $h_{n}=cn^{-1/8}$. 
``Max-$t$'' and ``$Wald$'' stand for, respectively, the Bonferroni-corrected Max-$t$-test, and Wald balancing tests based on $\widehat{ATE}_{n}\left( X^{j}\right)$ defined in (\ref{ate.x}).  ``$CvM_{n}^{unp}$'' and ``$KS_{n}^{unp}$'' are the  Cram\'{e}r-von Mises and Kolmogorov-Smirnov tests based on the unprojected empirical process $\hat{R}_{n}(u)$ defined in (\ref{R1}), whereas ``$CvM_{n}^{trad}$'' and ``$KS_{n}^{trad}$'' are defined analogously, but based on $\hat{R}_{n}^{trad}\left( x\right)$ as defined in (\ref{rtrad}). Finally, ``Bias'', ``CI length', and ``Coverage'' stands for the average simulated bias, estimated 95\% confidence interval length, and 95\% coverage probability for the $ATE$ estimator $\widehat{ATE_{n}}$ as defined in (\ref{ate}). All entries are proportions of rejections at 5\% level, in percentage points, except ``Bias'' ,``CI length'', and ``Coverage'' (measure in percentage points), which are as described above. See the main text for further details.}
\end{tablenotes}
\end{threeparttable}
\end{adjustbox}
\end{table}
\end{landscape}
}

Overall, the simulation results highlight that the proposed projection-based
tests perform favorably compared to other alternative testing procedures in
terms of size and power. Importantly, the simulations illustrate that the
gains in power can be credited to each distinguished feature of our tests,
that is, $\left( i\right) $ the avoidance of smoothing parameters by using
the ICM approach, $\left( ii\right) $ the dimension-reduction coming from
considering $1\left\{ q\left( X,\theta _{0}\right) \leq u\right\} $ instead
of $1\left\{ X\leq u\right\} $, and $\left( iii\right) $ the use of
orthogonal projections. Given these attractive features, we believe that our
tests can be of great use in practice.

\section{Empirical Illustration\label{application}}

In this section, we provide an empirical illustration of our testing
procedure. We revisit the analysis of \cite{Frankel2005} and \cite%
{Millimet2009}, and study the effect of trade on the environment. More
specifically, following \cite{Millimet2009}, we assess the effect of a
country being a member of the General Agreement on Tariffs and Trade (GATT)
or World Trade Organization (WTO) on 5 different measures of environmental
quality: per capita $CO_{2}$ emissions, average annual deforestation rate
for 1990-1996, energy depletion, rural access to clear water and urban
access to clear water.

As \cite{Millimet2009}, we use three different covariates to model the
probability of being a GATT/WTO member: real per capita GDP, land area per
capita, and polity, which is a measure of how democratic (versus autocratic)
is the structure of the government. The motivation to include these three
covariates is to increase the plausibility of Assumption \ref{unc} as
discussed by \cite{Frankel2005}. GDP per capita is associated with the
probability of being member of the GATT/WTO and at the same time may have
effects on different measures of environment quality, for instance, via the
environmental Kuznetz curve. Land area per capita is another potential
confounder since higher population density may lead to environmental
degradation and \textquotedblleft larger\textquotedblright\ countries are
more likely to trade more, affecting the probability of being member of he
GATT/WTO. Finally, as noted by \cite{Frankel2005}, low-democracy countries
tend to have lower measures of environmental quality, and can also
confound the effect of GATT/WTO membership. In what follows and in the
same spirit of \cite{Millimet2009}, we assume that Assumption \ref{unc}
holds after controlling for these three confounding factors\footnote{%
If one finds the plausibility of this assumption rather low, all the
estimates presented below should be interpreted as associations/correlations
and not as causal effects. In light of Remark \ref{rem:unc}, however,
Assumption \ref{unc} plays no role in our specification tests.}.

The unbalanced country-level panel data we use follows from \cite%
{Millimet2009}, and includes observations from 1990 (before the WTO) and
1995 (after the WTO). However, it is important to have in mind that
treatment is defined as being a GATT/WTO member, and therefore there are
countries who were treated in both times, and others who were treated only
in 1995. Table \ref{tab:sum} provides summary statistics and more detailed
description of the variables. Finally, we highlight that the data we analyze
is from an unbalanced country-level panel and instead of only considering
the \textquotedblleft always observed\textquotedblright\ countries, we
follow \cite{Millimet2009} and run a separate analysis for each outcome. For
further details, see Section 4.1 of \cite{Millimet2009}. \medskip

\begin{table}[h]
\caption{Summary statistics and variable descriptions}
\label{tab:sum}\centering
\begin{adjustbox}{ max width=1\linewidth, max totalheight=1\textheight, keepaspectratio}
\begin{threeparttable}
    \begin{tabular}{lcccl}
    \toprule
    Variable & Mean  & Standard deviation & $N$     & Description \\ \hline
    \midrule
    Per capita $CO_2$ & 3.82  & 4.73  & 232   & Carbon dioxide emissions, industrial, in metric tons per capita \\
    Deforestation & 0.68  & 1.28  & 223   & Annual deforestation, average percentage change, 1990-1995 \\
    Energy depletion & 3.13  & 7.43  & 223   & In percent of GDP, equal to the product of unit resource rents and the \\
          &       &       &       & physical quantities of fossil fuel energy extracted \\
    Rural water access & 51.20  & 27.42 & 137   & Access to clean water, percentage of rural population, 1990-1996 \\
    Urban water access & 76.28 & 21.76 & 140   & Access to clean water, percentage of urban population, 1990-1996 \\
    GATT/WTO membership & 0.78  & 0.41  & 232   & Member country of GATT/WTO: 1 if member, 0 otherwise \\
    Real GDP per capita & 7.30   & 7.47  & 232   & Real (1990) gross domestic product per capita, in thousands of dollars \\
    Polity & 3.17  & 6.85  & 232   & Index, ranging from-10 (strongly autocratic) to 10 (strongly democratic) \\
    Area per capita & 51.6  & 89.56 & 232   & Land area divided by population \\ \hline
    \bottomrule
    \end{tabular}%

    \begin{tablenotes}[para,flushleft]
\small{
Note: Same data used by \cite{Millimet2009}. We thank Prof. Millimet for sharing the data with us. Original source is Environmental indicators and country-level controls are from \cite{Frankel2005}, whereas GATT/WTO membership data are from \cite{Rose2004}. $N=$ number of observations. For further details, see the aforementioned papers.}
\end{tablenotes}
  
\end{threeparttable}
\end{adjustbox}
\end{table}

The main goal of this section is to assess the \textquotedblleft
reliability\textquotedblright\ of different treatment effect measures by
analyzing if different propensity score models are correctly specified or
not. Given that the sample constitutes of an unbalanced panel, we estimate
separate propensity score models for each outcome subsample. More
specifically, for each outcome, we model the probability of a country being
a GATT/WTO member ($D=1$ if member, $D=0$ otherwise) by a standard Probit
model and consider two different specifications:

\textit{Spec1:} $X$ includes real per capita GDP, land area per capita, and
polity.

\textit{Spec2:} $X$ is defined as in \textit{Spec1} but adds pairwise
interaction terms between each covariate.

For each of these specifications, we test the null hypothesis 
\begin{equation*}
H_{0}:\exists \theta _{0}\in \Theta :\mathbb{E}\left[ D-\Phi \left(
X^{\prime }\theta _{0}\right) |\Phi \left( X^{\prime }\theta _{0}\right) %
\right] =0~a.s.,
\end{equation*}%
against $H_{1}$, which is simply the negation of $H_{0}$. Table \ref%
{tab:tests} reports the testing results for each specification, together
with normalized IPW estimator for the $ATE$ based on (\ref{ate}). We also
consider the normalized IPW estimator for the $ATT$,%
\begin{equation}
\widehat{ATT}_{n}=\frac{1}{n}\sum_{i=1}^{n}\left( \frac{w_{1,i}^{treat}}{%
\bar{w}_{1,n}^{treat}}-\frac{w_{0,i}^{treat}}{\bar{w}_{0,n}^{treat}}\right)
Y_{i},  \label{ATT}
\end{equation}%
where $w_{1,i}^{treat}=D_{i}$, $w_{0,i}^{treat}=\left( 1-D_{i}\right)
q\left( X_{i},\hat{\theta}_{n}\right) /\left( 1-q\left( X_{i},\hat{\theta}%
_{n}\right) \right) $, and $\bar{w}_{d,n}^{treat}$ is the sample mean of $%
w_{d,i}^{treat},$ $d=\left\{ 0,1\right\} $. The associated standard errors
and $p$-values are in parenthesis and brackets, respectively. Following\ 
\cite{Millimet2009}, we trim observations with estimated propensity score
outside the interval $[0.05,0.95]$ to avoid denominators arbitrarily close
to zero. Bootstrapped $p$-values for our proposed specification tests are
based on 100,000 bootstrap draws\footnote{%
Note that the variables in \textit{Spec1 }and\textit{\ Spec2 }are all
functions of the same three covariates: real per capita GDP, land area per
capita, and polity. As so, \textit{Spec1 and Spec2} have the same
information content with respect to the reliability of Assumption \ref{unc}.}%
.

At the 5\% level we find that, based on the $CvM_{n}$ test statistic (\ref%
{cvm}), \textit{Spec1} is rejected for per capita $CO_{2},$ deforestation
and energy depletion, but is not rejected for rural and urban accesses to
clean water. The evidence of propensity score misspecification is weaker
when using the $KS_{n}$ test statistic (\ref{ks}). \textit{Spec2}, on the
other hand, is not rejected for any outcome at the usual significance
levels, using either $CvM_{n}$ or $KS_{n}$ test statistic. Thus, our tests
suggests that \textit{Spec2} should be preferred when analyzing per capita $%
CO_{2},$ deforestation and energy depletion, whereas for urban and rural
water access, our tests do not favor either specification.

\begin{table}[tbph]
\caption{Effect of GATT/WTO membership on environmental quality}
\label{tab:tests}\centering%
\begin{adjustbox}{ max width=1\linewidth, max totalheight=1\textheight, keepaspectratio}
\begin{threeparttable}
\begin{tabular}{lcccccccccccccc}
\toprule 
 & \multicolumn{2}{c}{Per capita $CO_{2}$} &  & 
\multicolumn{2}{c}{Deforestation} &  & \multicolumn{2}{c}{Energy Depletion}
&  & \multicolumn{2}{c}{Rural water access} &  & \multicolumn{2}{c}{Urban
water access} \\ \cline{2-3}\cline{5-6}\cline{8-9}\cline{11-12}\cline{14-15}
\cmidrule{2-3} \cmidrule{5-6} \cmidrule{8-9} \cmidrule{11-12} %
\cmidrule{14-15} & $Spec1$ & $Spec2$ &  & $Spec1$ & $Spec2$ &  & Spec 1 & $%
Spec2$ &  & $Spec1$ & $Spec2$ &  & $Spec1$ & $Spec2$ \\ 
\cmidrule{2-3} \cmidrule{5-6} \cmidrule{8-9} \cmidrule{11-12} %
\cmidrule{14-15} 

$\widehat{ATE_{n}}$& -1.29 & -1.00 &       & 0.26  & 0.34  &       & -3.35 & -3.39 &       & 3.07  & 2.89  &       & -5.21 & -4.62 \\
	    & (0.58)  & (0.49)  &       & (0.20)  & (0.21)  &       & (1.35)  & (1.38)  &       & (4.84)  & (4.79)  &       & (3.75)  & (3.72) \\
    &$[0.025]$ &$[0.039]$  &       &$[0.203]$  & $[0.105]$ &       & $[0.013]$  & $[0.014]$  &       & $[0.526]$ &$[0.547]$  &       &$ [0.165]$  &$ [0.213]$ \\
     &   &   &       &  &   &       &  &   &       &   & &       &   &  \\
$\widehat{ATT_{n}}$	 & -0.81 & -0.56 &       & 0.14  & 0.22  &       & -2.16 & -1.62 &       & 1.74  & 1.33  &       & 1.74  & -3.34 \\
     & (0.68)  & (0.59)  &       & (0.20)  & (0.21)  &       & (1.30)  & (1.34)  &       & (5.02)  & (4.98)  &       & (4.19)  & (4.16)  \\ 
     &$[0.234]$ &$[0.338]$  &       &$[0.500]$  & $[0.287]$ &       & $[0.097]$  & $[0.228]$  &       & $[0.730]$ &$[0.790]$  &       &$ [0.679]$  &$ [0.422]$ \\
     &   &   &       &  &   &       &  &   &       &   & &       &   &  \\
$CvM_{n}$ & 0.01  & 0.28  &       & 0.01  & 0.45  &       & 0.01  & 0.35  &       & 0.53  & 0.26  &       & 0.49  & 0.23 \\
$KS_{n}$ & 0.12  & 0.24  &       & 0.17  & 0.57  &       & 0.12  & 0.43  &       & 0.70  & 0.52  &       & 0.63  & 0.28 \\
\bottomrule
\end{tabular}
\begin{tablenotes}[para,flushleft]
\small{
Note: $Spec1$ and $Spec2$ are different specifications of the propensity score.
$\widehat{ATE_{n}}$ and  $\widehat{ATT_{n}}$ are the estimators for $ATE$ and $ATT$ in (\ref%
{ate}) and (\ref%
{ATT}), respectively, but with observations with estimated propensity score outside $[0.05,0.95]$ trimmed. Standard errors are in parenthesis, and $p$-values in brackets. 
 ``$CvM_{n}$'' and ``$KS_{n}$'' respectively stand for the bootstrapped $p$-values of our proposed Cram\'{e}r-von Mises and Kolmogorov-Smirnov tests based on 100,000 bootstrap draws. See the main text for further details.}
\end{tablenotes}
\end{threeparttable}
\end{adjustbox}
\end{table}

Next we comment on the consequences of propensity score misspecification.
For per capita $CO_{2}$, our results suggest that the overall effect of
GATT/WTO membership on emissions is negative and statistically significant
at the 5\% level under both propensity score specifications. On the other
hand, we find the effect of GATT/WTO membership on per capita $CO_{2}$ is
not statistically significant among the treated sub-population using either
specification. In terms of point estimates, however, there are important
differences. For example, the $ATE$ point estimate under \textit{Spec1}
(misspecified propensity score) is 30\% higher (in absolute terms) than
under \textit{Spec2}. Note that the $0.3$ difference in $ATE$ represents
roughly 8\% of the overall per capita $CO_{2}$ emissions.

When we analyze the effect of GATT/WTO on deforestation and energy
depletion, our results again highlight the consequences of propensity score
misspecifications. We find that the $ATE$ point estimate for the effect of
GATT/WTO membership on deforestation is 30\% larger under \textit{Spec2}
than under \textit{Spec1}, and the $ATT$ point estimate for the effect of
GATT/WTO membership on energy depletion is 25\% smaller under \textit{Spec2 }%
than under \textit{Spec1}. Such large differences are economically
significant, as the 0.08 difference in $ATEs$ on deforestation represents
nearly 12\% of the mean annual deforestation, and the 0.46 difference in $%
ATTs$ on energy depletion represents nearly 15\% of the mean energy
depletion among countries in the sample. Interestingly, the $ATT$ on energy
depletion is statistically significant at 10\% level under \textit{Spec1},
but not under \textit{Spec2}, highlighting that propensity score
misspecifications can also lead to invalid inference. Finally, we note that
although our tests do not detect propensity score misspecification for the
rural and urban water access, the results suggest that GATT/WTO membership
is not statistically significant using either propensity score
specification, perhaps because of the relatively high standard errors due to
the limited sample size.

Overall, we find that propensity score misspecifications can affect both the
economic and statistical conclusions about the effect of GATT/WTO membership
on environmental quality. When the propensity score is correctly specified,
our results suggest that GATT/WTO membership is associated with improved
environmental performance in terms of $CO_{2}$ and energy depletion but not
in terms of rural and urban water access. We also find that GATT/WTO
membership is associated with higher deforestation, though the statistical
evidence is relatively weaker. Furthermore, our results uncover interesting
heterogeneity, as the aforementioned results are statistically significant
for the overall population ($ATE$) but not for the treated-subpopulation ($%
ATT$).

\section{Conclusion\label{conclusion}}

In this article, we have shown that, when propensity scores are correctly
specified, a particular restriction between the propensity score CDFs of
treated and control groups must hold. Based on such restriction, we propose
new nonparametric projection-based tests for the correct specification of
the propensity score. In contrast to other proposals, our tests are not
severely affected by the \textquotedblleft curse of
dimensionality\textquotedblright ', are not sensitive to the different
estimation methods used to estimate the propensity score under the null, do
not rely on the potentially ad hoc choice of bandwidths, and enjoy some
optimal power properties against particular classes of alternative
hypotheses. We have derived the asymptotic properties of the proposed tests,
and have proved that they are able to detect local alternatives converging
to the null at the parametric rate, and that critical values can be easily
computed via a simple multiplier bootstrap procedure. Our Monte Carlo
simulation study illustrates that, for a large class of alternatives, our
projection-based tests perform better in finite samples than existing tests,
though there are some classes of alternatives in which our tests have
trivial power. All these finite sample findings are in line with our
asymptotic results. Finally, our empirical application concerning the effect
of trade on the environment shows the feasibility and appeal of our tests in
relevant scenarios. Given that the validity of many policy evaluation
procedures relies on the correct specification of the propensity score, we
argue that the tests proposed in this article are important additions to the
applied researchers' toolkit.

We would like to mention that, in general, our tests should be seen as a
\textquotedblleft model validation\textquotedblright\ and not a
\textquotedblleft model selection\textquotedblright\ procedure. Once a
propensity score model is selected, our specification tests can provide
evidence of its reliability or lack thereof. In case the putative propensity
score model is rejected, one can consider more flexible specifications.
For instance, one can add additional interaction terms into the original model, consider semiparametric single-index or partially linear models, among
other possible strategies. Having said so, we emphasize that if one uses our
testing procedure as a \textquotedblleft model selection\textquotedblright\
device, one must bear in mind that standard inference procedures for
treatment effects may be invalid if one treats the resulting selected
propensity score as the \textquotedblleft true\textquotedblright\ one, see
e.g. \cite{Leeb2005}. Thus, in case one uses our proposal for model
selection, one must account for the model selection step in order to make
valid inference about the treatment effect, see e.g. \cite{Belloni2014}, 
\cite{Chernozhukov2016}, and \cite{Belloni2017}. A full discussion of this
procedure is beyond the scope of this article and we leave it for future
research.

Finally, we note that results in Lemma \ref{lem1} can also be used for
estimating the propensity score such that, for a given specification $%
q\left( X,\theta _{0}\right) $, condition $\mathbb{E}\left[ D-q\left(
X,\theta _{0}\right) |q\left( X,\theta _{0}\right) \right] =0~a.s.$ is
directly imposed. One could use the minimum distance method described in 
\cite{Dominguez2004} to estimate $\theta _{0}$, for example. A detailed
discussion of this estimator is beyond the scope of this article and is
deferred to future work.

\begin{appendices}

\section*{\center{Appendix A: Mathematical Proofs}}\label{appendix_a}

\renewcommand{\thesection}{A} \renewcommand{\theassumption}{A.%
\arabic{assumption}} \renewcommand{\thelemma}{A.\arabic{lemma}} %
\setcounter{lemma}{0} \setcounter{equation}{0} \setcounter{figure}{0} %
\setcounter{section}{0}

We provide the proofs of our main theoretical results in this appendix. We
first prove Lemma \ref{lem1}.

\noindent \textbf{Proof of Lemma \ref{lem1}:} To begin, let $F\left(
u\right) =\mathbb{P}\left( p\left( X\right) \leq u\right) $, and for $%
d=\left\{ 0,1\right\} $, $F_{d}\left( u\right) =\mathbb{P}\left( p\left(
X\right) \leq u|D=d\right) .$ Denote the density of $F\left( u\right) $, $%
F_{1}\left( u\right) $ and $F_{0}\left( u\right) $ by $f\left( u\right) $, $%
f_{1}\left( u\right) $, and $f_{0}\left( u\right) $. The proof of (\ref{cs1}%
) follows from Lemma 3.1 in \cite{Shaikh2009}, as they proved that, for all $%
0<u<1$ inside the support of the propensity score $p\left( X\right) $, 
\begin{equation}
f_{1}\left( u\right) =\alpha \frac{u}{1-u}f_{0}\left( u\right) .
\label{lem1eq}
\end{equation}%
Thus, (\ref{cs1}) follows from integrating both sides of (\ref{lem1eq}).

Next, we prove (\ref{ort1}). By straightforward manipulation of (\ref{cs1}),
we have that, for all $u\in \left( 0,1\right) $, 
\begin{eqnarray}
\mathbb{E}\left[ 1\left\{ p\left( X\right) \leq u\right\} |D=1\right] 
\mathbb{P}\left( D=1\right) &=&\mathbb{P}\left( D=0\right) ~\mathbb{E}\left[ 
\frac{p\left( X\right) }{1-p\left( X\right) }1\left\{ p\left( X\right) \leq
u\right\} |D=0\right]  \notag \\
\mathbb{E}\left[ D1\left\{ p\left( X\right) \leq u\right\} \right] &=&%
\mathbb{E}\left[ \left( 1-D\right) \frac{p\left( X\right) }{1-p\left(
X\right) }1\left\{ p\left( X\right) \leq u\right\} \right]  \notag \\
\mathbb{E}\left[ \frac{\left( D-p\left( X\right) \right) }{1-p\left(
X\right) }1\left\{ p\left( X\right) \leq u\right\} \right] &=&0\mathbb{.~}
\end{eqnarray}%
Note that 
\begin{eqnarray}
\mathbb{E}\left[ \frac{\left( D-p\left( X\right) \right) }{1-p\left(
X\right) }1\left\{ p\left( X\right) \leq u\right\} \right] &=&0~a.e.\text{
in }u\in \left( 0,1\right)  \notag \\
&\iff &  \notag \\
\mathbb{E}\left[\left. \frac{\left( D-p\left( X\right) \right) }{1-p\left(
X\right) }\right\vert p\left( X\right) \right] &=&0~a.s.,  \label{lem1eq2}
\end{eqnarray}%
see e.g. Lemma 1 in \cite{Escanciano2006a}. Given that $p\left( X\right) $
is bounded away from one, we have that (\ref{lem1eq2}) is equivalent to (\ref%
{ort1}). Finally, because $0<p\left( X\right) <1$ $a.s.$, we can trivially
include the two boundary points $u=0$ and $u=1$, concluding our proof. $\square $%
\bigskip

Next, we state and prove several auxiliary lemmas that help us prove our
main theorems. Let us first introduce some notation. Let $Y$ be a generic
random variable with cumulative distribution function $F$. Recall the
definition of the quantile function $F^{-1}$ associated with $F$, namely, $%
F^{-1}\left( u\right) =\inf \left\{ y\in \mathbb{R}:F\left( y\right) \geq
u\right\} $ for $0\leq u\leq 1$, and let $F\left( a-\right) \equiv \lim_{y\uparrow
a}F\left( y\right) .$

\begin{lemma}
\label{lemPIT} Let $A=\left\{ a\right\} $ denote the set of atoms of $F$,
and let $V$ be independent of $Y$ and uniformly distributed on $\left[ 0,1%
\right] $. Set 
\begin{equation*}
U=\left\{ 
\begin{array}{ll}
F\left( Y\right) , & \phantom{-----}if~Y\not\in A, \\ 
F\left( a-\right) +F\left\{ a\right\} V, & \phantom{-----}if~Y=a~for~a\in A,%
\end{array}%
\right.
\end{equation*}%
where $F\left\{ a\right\} \equiv F\left( a\right) -F\left( a-\right) $.
Then, it follows that\newline
(i) $U$ is uniformly distributed on $\left[ 0,1\right] $;\newline
(ii) $Y=F^{-1}\left( U\right) $ $a.s.$;\newline
(iii) $1\left\{ Y\leq y\right\} =1\left\{ U\leq F(y)\right\} $ $a.s.$ for
all $y\in \mathbb{R}$.
\end{lemma}

\textbf{Proof of Lemma \ref{lemPIT}:} This lemma follows directly from
Proposition 3.2 in Chapter 7 of \cite{Shorack2000}; see also Lemma 2.8 in 
\cite{Stute1993a} for an earlier use of such quantile transformation. For
completeness, we give a short proof of part $\left( i\right) $ below. Part $%
\left( ii\right) $ follows directly from the construction of $U$ and the
definition of the quantile function, and part $(iii)$ is a consequence of
parts $(i)$ and $\left( ii\right) $.

Without loss of generality, let us denote $A=\left\{ a_{1},a_{2},\ldots
\right\} $, with $-\infty \leq a_{1}<a_{2}<\cdots <a_{J}\leq \infty $ and $%
\mathbb{P}\left( Y=a_{l}\right) =F\left\{ a_{l}\right\} =F\left(
a_{l}\right) -F\left( a_{l}-\right) >0$ for $l=1,2,\ldots ,J$ (note that $J$
can be $\infty $). Without loss of generality, assume that $%
0<\sum_{l=1}^{J}F\left\{ a_{l}\right\} <1$, so $Y$ is not purely discrete. The case of $\sum_{l=1}^{J}F\left\{ a_{l}\right\}=0$ or $1$ is trivial.

For $u\in \left[ 0,1\right] $, the CDF of the distribution of $U$ is 
\begin{equation*}
\mathbb{P}\left( U\leq u\right) =\mathbb{P}\left( U\leq u,Y\not\in A\right)
+\sum_{l}\mathbb{P}\left( U\leq u|Y=a_{l}\right) \mathbb{P}\left(
Y=a_{l}\right) ,
\end{equation*}%
with 
\begin{align*}
\mathbb{P}\left( U\leq u|Y=a_{l}\right) =& \mathbb{P}\left( V\leq \frac{%
u-F(a_{l}-)}{F\left\{ a_{l}\right\} }\right) \\
=& \frac{u-F(a_{l}-)}{F\left\{ a_{l}\right\} }1\left\{ F\left( a_{l}-\right)
\leq u<F\left( a_{l}\right) \right\} +1\left\{ F\left( a_{l}\right) \leq
u\leq 1\right\} .
\end{align*}%
We obtain 
\begin{align*}
\mathbb{P}\left( U\leq u\right) =& \mathbb{P}\left( U\leq u,Y\not\in A\right)
\\
& +\sum_{l}\left[ \left( u-F(a_{l}-)\right) 1\left\{ F\left( a_{l}-\right)
\leq u<F\left( a_{l}\right) \right\} +F\left\{ a_{l}\right\} 1\left\{
F\left( a_{l}\right) \leq u\leq 1\right\} \right] .
\end{align*}

Four cases then stand out. Let $j$ be some generic integer such that $2\leq
j\leq J-1$.

Case 1: If $u\in[F(a_j-),F(a_j))$, 
\begin{equation*}
\mathbb{P}\left(U\leq u,Y\not\in A\right)=\mathbb{P}\left(Y\leq
F^{-1}(u),Y\not\in A\right)=F(a_j-)-F\{a_{j-1}\}-\ldots-F\{a_1\},
\end{equation*}
and, noting that the intervals $[F(a_l),1]$ for $1\leq l\leq j-1$ always
contain the interval $[F(a_j),1]$, 
\begin{align*}
&\sum_l\left[\left(u-F(a_l-)\right)1\left\{F\left(a_l-\right)\leq
u<F\left(a_l\right)\right\}+F\left\{a_l\right\}1\left\{F\left(a_l\right)\leq
u\leq 1\right\}\right] \\
=&\left(u-F(a_j-)\right)+\left(F\{a_1\}+\ldots+F\{a_{j-1}\}\right).
\end{align*}
As a result, we find $\mathbb{P}\left(U\leq u\right)=u$ when $%
u\in[F(a_j-),F(a_j))$.

Case 2: If $u\in[F(a_{j-1}),F(a_{j}-))$, similar arguments as in Case 1
yields 
\begin{align*}
\mathbb{P}\left(U\leq
u\right)=\left(u-F\{a_{j-1}\}-\ldots-F\{a_{1}\}\right)+\left(F\{a_1\}+%
\ldots+F\{a_{j-1}\}\right)=u.
\end{align*}

Case 3: If $u\in[0,F(a_1))$, obviously, 
\begin{align*}
\mathbb{P}\left(U\leq u\right)=u.
\end{align*}

Case 4: If $u\in[F(a_{J}),1]$, obviously, 
\begin{align*}
\mathbb{P}\left(U\leq
u\right)=\left(u-F\{a_{J}\}-\ldots-F\{a_{1}\}\right)+\left(F\{a_1\}+\ldots+F%
\{a_{J}\}\right)=u.
\end{align*}

Thus, it follows that $U$ is uniformly distributed on $\left[ 0,1\right] $. $%
\square $\bigskip

\begin{lemma}
\label{lemDonsker} Under Assumptions \ref{ass1} and \ref{ass3}, $\mathcal{F}%
=\left\{x\mapsto 1\left\{q(x,\theta )\leq u\right\}:u\in \left[ 0,1\right], \theta \in \Theta
\right\}$ is a Donsker class of functions.
\end{lemma}

\noindent \textbf{Proof of Lemma \ref{lemDonsker}: }\ From Lemma \ref{lemPIT}, simply letting $Y=q(X,\theta )$, we have that $1\left\{ q(X,\theta )\leq
u\right\} =1\left\{ U\leq F_{\theta }\left( u\right) \right\} $ for all $%
u\in \left[ 0,1\right] $, i.e., we can exploit the quantile transformation
and express each indicator function $1\left\{ q(X,\theta )\leq u\right\} $
through $U$ via the time transformation $F_{\theta }$. In light of this
quantile transformation, it suffices to study the class of functions $%
\mathcal{F}_{cdf}=\left\{ \bar{u}\mapsto 1\left\{ \bar{u}\leq F_{\theta
}\left( u\right) \right\} :u\in \left[ 0,1\right] ,\theta \in \Theta
\right\} .$

Let $N_{[\,\,]}(\bar{\epsilon},\mathcal{F}_{cdf},\mathcal{L}_{2}(\mathbb{P}%
)) $ be the bracketing number of the class $\mathcal{F}_{cdf}$ with respect
to the underlying probability $\mathbb{P}$, which by definition is the
minimal number of $\bar{\epsilon}$-brackets under $\mathcal{L}_{2}(\mathbb{P}%
)$ metric that cover $\mathcal{F}_{cdf}$. Henceforth, we define the
underlying probability $\mathbb{P}$ as the probability measure of $U$. By
Theorem 2.5.6 in \cite{VanderVaart1996}, the Donsker property is implied by 
\begin{equation*}
\int_{0}^{\infty }\sqrt{\log N_{[\,\,]}(\bar{\epsilon},\mathcal{F}_{cdf},\mathcal{L%
}_{2}(\mathbb{P}))}\,d\bar{\epsilon}<\infty .
\end{equation*}%
To show that such entropy result hold, we follow similar steps as the proof
of Lemma 1 in \cite{Akritas2001} and of Lemma A.4 in \cite{Frazier2018}.

Let $\bar{\epsilon}>0$ be an arbitrarily small constant, and consider
partitions $\left\{ \Theta _{l}\right\} _{l=1}^{L}$ of $\Theta $. Given that 
$\Theta $ is compact, under Assumption \ref{ass3}, there exists a finite
constant $K\leq diam\left( \Theta /\bar{\epsilon}\right) ^{k}$ such that $%
diam\left( \Theta _{l}\right) \leq \bar{\epsilon}$ for every $l=1,\dots ,K$.
Fix $l\in \left\{ 1,\dots ,K\right\} $ and pick up some $\theta _{l}\in
\Theta _{l}$. Then, for any fixed $u\in \left[ 0,1\right] $ and any $\theta
\in \Theta _{l}$, it follows from the Lipschitz condition in Assumption \ref%
{ass3} that%
\begin{equation}
F_{l}^{-}\left( u\right) \leq F_{\theta }\left( u\right) \leq
F_{l}^{+}\left( u\right) ,  \label{lipschitz}
\end{equation}%
where $F_{l}^{\pm }(u)\equiv F_{\theta _{l}}\left( u\right) \pm \bar{\epsilon%
}C$, where $C$ is as defined in Assumption \ref{ass3}. Thus, for each $%
\theta \in \Theta _{l}$ and each $u\in \left[ 0,1\right] $, it follows that 
\begin{equation*}
1\left\{ \bar{u}\leq F_{l}^{-}\left( u\right) \right\} \leq 1\left\{ \bar{u}%
\leq F_{\theta }\left( u\right) \right\} \leq 1\left\{ \bar{u}\leq
F_{l}^{+}\left( u\right) \right\} .
\end{equation*}

Denote $F_{l}^{L}\left( u\right) =\mathbb{P}\left( U\leq F_{l}^{-}\left(
u\right) \right) $ and let $u_{l,j}^{L}$, $j=1,\dots ,K\bar{\epsilon}^{-1}$,
partition the unit interval in segments such that $F_{l}^{L}\left(
u_{l,j}^{L}-\right) -F_{l}^{L}\left( u_{l,j-1}^{L}\right) <\bar{\epsilon}$.
\ Similarly, define $F_{l}^{U}\left( u\right) =\mathbb{P}\left( U\leq
F_{l}^{+}\left( u\right) \right) $ and let $u_{l,j}^{U}$, $j=1,\dots ,K\bar{%
\epsilon}^{-1}$, partition the unit interval in segments such that $%
F_{l}^{U}\left( u_{l,j}^{U}-\right) -F_{l}^{U}\left( u_{l,j-1}^{U}\right) <%
\bar{\epsilon}$. Now, define the following bracket for $u:$%
\begin{equation*}
u_{l,j_{1}}^{L}\leq u\leq u_{l,j_{2}}^{U},
\end{equation*}%
where $u_{l,j_{1}}^{L}$ is the largest of the $u_{l,j}^{L}$ with the
property of being less than or equal to $u$ and $u_{l,j_{2}}^{U}$ is the
smallest $u_{l,j}^{U}$ among those that are greater than or equal to $u$.
Obviously, we have 
\begin{equation*}
1\left\{ \bar{u}\leq F_{l}^{-}\left( u_{l,j_{1}}^{L}\right) \right\} \leq
1\left\{ \bar{u}\leq F_{\theta }\left( u\right) \right\} \leq 1\left\{ \bar{u%
}\leq F_{l}^{+}\left( u_{l,j_{2}}^{U}\right) \right\} .
\end{equation*}%
Then, for an arbitrary constant $K$, 
\begin{eqnarray*}
&&\left\Vert 1\left\{ U\leq F_{l}^{+}\left( u_{l,j_{2}}^{U}\right) \right\}
-1\left\{ U\leq F_{l}^{-}\left( u_{l,j_{1}}^{L}\right) \right\} \right\Vert
_{2}^{2} \\
&=&\mathbb{E}\left[ 1\left\{ U\leq F_{l}^{+}\left( u_{l,j_{2}}^{U}\right)
\right\} -1\left\{ U\leq F_{l}^{-}\left( u_{l,j_{1}}^{L}\right) \right\} %
\right] \\
&\leq &F_{l}^{+}\left( u_{l,j_{2}}^{U}\right) -F_{l}^{-}\left(
u_{l,j_{1}}^{L}\right) \\
&=&\left( F_{l}^{+}\left( u\right) -F_{l}^{-}\left( u\right) \right) +\left(
F_{l}^{+}\left( u_{l,j_{2}}^{U}\right) -F_{l}^{+}\left( u\right) \right)
+\left( F_{l}^{-}\left( u\right) -F_{l}^{-}\left( u_{l,j_{1}}^{L}\right)
\right) \\
&\leq &K\bar{\epsilon},
\end{eqnarray*}%
where the last inequality follows from (\ref{lipschitz}) and the definition
of definitions of $u_{l,j_{2}}^{U}$ and $u_{l,j_{1}}^{L}$. Consequently, we
have $\left\Vert 1\left\{ U\leq F_{l}^{+}\left( u_{l,j_{2}}^{U}\right)
\right\} -1\left\{ U\leq F_{l}^{-}\left( u_{l,j_{1}}^{L}\right) \right\}
\right\Vert _{2}\leq K\bar{\epsilon}^{1/2}$.

Thus, the bracketing number $N_{[\,\,]}(\bar{\epsilon},\mathcal{F}_{cdf},%
\mathcal{L}_{2}(\mathbb{P}))$ is of polynomial order $\left( 1/\bar{\epsilon}%
\right) $, and the entropy is of smaller order than $\log \left( 1/\bar{%
\epsilon}\right) $. Therefore, since $\mathcal{F}_{cdf}$ is uniformly
bounded between 0 and 1, we conclude that 
\begin{equation*}
\int_{0}^{\infty }\sqrt{\log N_{[\,\,]}(\bar{\epsilon},\mathcal{F}_{cdf},%
\mathcal{L}_{2}(\mathbb{P}))}\,d\bar{\epsilon}\leq K\int_{0}^{1}\sqrt{\log 
\bar{\epsilon}^{-1}}\,d\bar{\epsilon}<\infty ,
\end{equation*}%
with $\int_{0}^{1}\sqrt{\log \bar{\epsilon}^{-1}}\,d\bar{\epsilon}%
=-\int_{0}^{\infty }t\,de^{-t^{2}}=\sqrt{\pi }/2$, which implies that class $%
\mathcal{F}_{cdf}$ (and therefore class $\mathcal{F}$) is Donsker. $\square $

\bigskip

Define the auxiliary empirical process 
\begin{equation*}
\tilde{R}_{n}(u)=\frac{1}{\sqrt{n}}\sum_{i=1}^{n}\varepsilon _{i}(\hat{\theta%
}_{n})1\left\{ q(X_{i},\theta _{0})\leq u\right\} .
\end{equation*}%
Next lemma states that the unprojected process $\hat{R}_{n}(u)$ defined in (%
\ref{R1}) is asymptotically equivalent under $H_{0}$ to the auxiliary
process $\tilde{R}_{n}(u)$ defined above.

\begin{lemma}
\label{lemmaA2} Under Assumptions \ref{ass1}-\ref{ass3} and under the null
hypothesis $H_{0}$, we have 
\begin{equation*}
\sup_{u\in \Pi }\left\vert \hat{R}_{n}(u)-\tilde{R}_{n}(u)\right\vert
=o_{p}(1).
\end{equation*}
\end{lemma}

\noindent \textbf{Proof of Lemma \ref{lemmaA2}:} Note that we have uniformly
in $u$, 
\begin{align*}
& \hat{R}_{n}(u)-\tilde{R}_{n}(u) \\
=& \frac{1}{\sqrt{n}}\sum_{i=1}^{n}\varepsilon _{i}(\hat{\theta}%
_{n})\left(1\left\{ q(X_{i},\hat{\theta}_{n})\leq u\right\} -1\left\{
q(X_{i},\theta _{0})\leq u\right\}\right) \\
=& \frac{1}{\sqrt{n}}\sum_{i=1}^{n}(\varepsilon _{i}(\theta _{0})-(q(X_{i},%
\hat{\theta}_{n})-q(X_{i},\theta _{0})))\left(1\left\{ q(X_{i},\hat{\theta}%
_{n})\leq u\right\} -1\left\{ q(X_{i},\theta _{0})\leq u\right\}\right) \\
=& \frac{1}{\sqrt{n}}\sum_{i=1}^{n}\varepsilon _{i}(\theta
_{0})\left(1\left\{ q(X_{i},\hat{\theta}_{n})\leq u\right\} -1\left\{
q(X_{i},\theta _{0})\leq u\right\}\right) \\
&-\sqrt{n}(\hat{\theta}_{n}-\theta _{0})^{\prime }\frac{1}{n}%
\sum_{i=1}^{n}g(X_{i},\theta _{0})\left(1\left\{ q(X_{i},\hat{\theta}%
_{n})\leq u\right\} -1\left\{ q(X_{i},\theta _{0})\leq
u\right\}\right)+o_{p}(1) \\
:=& A_{1n}(u)+A_{2n}(u)+o_{p}(1),
\end{align*}%
where the second to last equality follows by the Taylor
expansion of $q(X_i,\hat{\theta}_n)$ around $q(X_i,\theta_0)$ and the Assumptions \ref{ass1} and \ref%
{ass2}. Following the arguments in the proof of Theorem 1 in \cite{Stute2002}%
, we next show that both $A_{1n}(u)$ and $A_{2n}(u)$ are uniformly
negligible in $u$.

For the first term $A_{1n}(u)$, define the process 
\begin{equation*}
\alpha _{n}(u,\theta )=\frac{1}{\sqrt{n}}\sum_{i=1}^{n}\varepsilon
_{i}(\theta _{0})1\left\{ q(X_{i},\theta )\leq u\right\} .
\end{equation*}%
Since under $H_{0}$ the $\varepsilon_i$'s are centered conditionally on $X_i$'s, $\alpha _{n}(u,\theta )$ has $i.i.d.$ centered summands. Clearly, the
first term $A_{1n}(u)$ can be expressed as $\alpha _{n}(u,\hat{\theta}%
_{n})-\alpha _{n}(u,\theta _{0})$. From Lemma \ref{lemDonsker}, $\alpha
_{n}(\cdot ,\cdot )$ is asymptotically equicontinuous, see e.g. Corollary
2.3.12 in \cite{VanderVaart1996}. Since $\hat{\theta}_{n}\rightarrow
_{p}\theta _{0}$ by Assumption \ref{ass1}, $A_{1n}(u)\rightarrow 0$ in
probability uniformly in $u$.

For the second term $A_{2n}(u)$, since by Assumption \ref{ass1}, $\sqrt{n}(%
\hat{\theta}_{n}-\theta _{0})=O_{p}(1)$, it remains to show that, uniformly
in $u$, 
\begin{equation*}
\frac{1}{n}\sum_{i=1}^{n}g(X_{i},\theta _{0})\left(1\left\{ q(X_{i},\hat{%
\theta}_{n})\leq u\right\} -1\left\{ q(X_{i},\theta _{0})\leq u\right\}
\right)\rightarrow _{p}0.
\end{equation*}%
However, this follows straightforwardly from the uniform convergence of 
\begin{equation*}
\frac{1}{n}\sum_{i=1}^{n}g(X_{i},\theta _{0})1\left\{ q(X_{i},\theta )\leq
u\right\}
\end{equation*}%
in $u$ and $\theta $ together with the continuity of its limit. $\square
\bigskip $

With the help of Lemma \ref{lemmaA2}, the next lemma establishes the
asymptotic uniform decomposition of the unprojected process $\hat{R}_{n}(u)$ in (\ref{R1}).

\begin{lemma}
\label{lemmaA3} Under Assumptions \ref{ass1}-\ref{ass3} and under the null
hypothesis $H_{0}$, we have 
\begin{equation*}
\sup_{u\in \Pi }\left\vert \hat{R}_{n}(u)-\frac{1}{\sqrt{n}}%
\sum_{i=1}^{n}\varepsilon _{i}(\theta _{0})1\left\{ q(X_{i},\theta _{0})\leq
u\right\} +\sqrt{n}(\hat{\theta}_{n}-\theta _{0})^{\prime }G(u,\theta
_{0})\right\vert =o_{p}(1),
\end{equation*}%
where $G(u,\theta )=\mathbb{E}\left[g(X,\theta )1\left\{ q(X,\theta )\leq
u\right\}\right]$.
\end{lemma}

\noindent \textbf{Proof of Lemma \ref{lemmaA3}:} From Lemma \ref{lemmaA2},
we immediately have uniformly in $u$, 
\begin{align*}
& \hat{R}_{n}(u)=\tilde{R}_{n}(u)+o_{p}(1) \\
=& \frac{1}{\sqrt{n}}\sum_{i=1}^{n}\varepsilon _{i}(\theta _{0})1\left\{
q(X_{i},\theta _{0})\leq u\right\} -\frac{1}{\sqrt{n}}\sum_{i=1}^{n}(q(X_{i},%
\hat{\theta}_{n})-q(X_{i},\theta _{0}))1\left\{ q(X_{i},\theta _{0})\leq
u\right\} +o_{p}(1).
\end{align*}%
By the Mean Value Theorem (MVT) and Assumption \ref{ass1}, the second term
in the previous expression is simply 
\begin{align*}
& -\sqrt{n}(\hat{\theta}_{n}-\theta _{0})^{\prime }\frac{1}{n}\sum_{i=1}^{n}%
\frac{\partial q(X_{i},\tilde{\theta}_{n})}{\partial \theta }1\left\{
q(X_{i},\theta _{0})\leq u\right\} \\
=& -\sqrt{n}(\hat{\theta}_{n}-\theta _{0})^{\prime }\mathbb{E}\left[%
g(X,\theta_{0})1\left\{ q(X,\theta _{0})\leq u\right\}\right]+o_{p}(1),
\end{align*}%
with $\tilde{\theta}_{n}$ lying between $\hat{\theta}_{n}$ and $\theta _{0}$%
, where the latter equality follows by the uniform law of large numbers
(ULLN) of \cite{Newey1994c}, Lemma 2.4. This finishes the proof of Lemma \ref%
{lemmaA3}. $\square \medskip $

Define the following quantity 
\begin{equation*}
\hat{S}_{n}=\frac{1}{\sqrt{n}}\sum_{i=1}^{n}\varepsilon _{i}(\hat{\theta}%
_{n})g(X_{i},\hat{\theta}_{n}).
\end{equation*}

\begin{lemma}
\label{lemmaA4}Under Assumptions \ref{ass1}-\ref{ass3} and under the null
hypothesis $H_{0}$, we have 
\begin{equation*}
\hat{S}_{n}=\frac{1}{\sqrt{n}}\sum_{i=1}^{n}\varepsilon _{i}(\theta
_{0})g(X_{i},\theta _{0})-\Delta (\theta _{0})\sqrt{n}(\hat{\theta}%
_{n}-\theta _{0})+o_{p}(1),
\end{equation*}%
where $\Delta (\theta )=\mathbb{E}[g(X,\theta )g^{\prime }(X,\theta )]$.
\end{lemma}

\noindent \textbf{Proof of Lemma \ref{lemmaA4}:} We can rewrite 
\begin{align*}
\hat{S}_{n}=& \frac{1}{\sqrt{n}}\sum_{i=1}^{n}\varepsilon _{i}(\theta
_{0})g(X_{i},\theta _{0})+\frac{1}{\sqrt{n}}\sum_{i=1}^{n}(\varepsilon _{i}(%
\hat{\theta}_{n})-\varepsilon _{i}(\theta _{0}))g(X_{i},\theta _{0}) \\
& +\frac{1}{\sqrt{n}}\sum_{i=1}^{n}\varepsilon _{i}(\theta _{0})(g(X_{i},%
\hat{\theta}_{n})-g(X_{i},\theta _{0}))+\frac{1}{\sqrt{n}}%
\sum_{i=1}^{n}(\varepsilon _{i}(\hat{\theta}_{n})-\varepsilon _{i}(\theta
_{0}))(g(X_{i},\hat{\theta}_{n})-g(X_{i},\theta _{0})) \\
:=& \frac{1}{\sqrt{n}}\sum_{i=1}^{n}\varepsilon _{i}(\theta
_{0})g(X_{i},\theta _{0})+C_{1n}+C_{2n}+C_{3n}.
\end{align*}

We first show that $C_{1n}=-\sqrt{n}(\hat{\theta}_{n}-\theta _{0})^{\prime
}\Delta (\theta _{0})+o_{p}(1)$. Note that 
\begin{align*}
C_{1n}=& -\frac{1}{\sqrt{n}}\sum_{i=1}^{n}(q(X_{i},\hat{\theta}%
_{n})-q(X_{i},\theta _{0}))g(X_{i},\theta _{0}) \\
=& -\frac{1}{n}\sum_{i=1}^{n}g(X_{i},\theta _{0})\frac{\partial q(X_{i},%
\tilde{\theta}_{n})}{\partial \theta ^{\prime }}\sqrt{n}(\hat{\theta}%
_{n}-\theta _{0}) \\
=& -\mathbb{E}[g(X,\theta _{0})g^{\prime }(X,\theta _{0})]\sqrt{n}(\hat{%
\theta}_{n}-\theta _{0})+o_{p}(1),
\end{align*}%
with $\tilde{\theta}_{n}$ lying between $\hat{\theta}_{n}$ and $\theta _{0}$%
, where the second equality follows by the MVT, and the last
equality follows from the ULLN of \cite{Newey1994c}, Lemma 2.4, and Assumptions \ref{ass1} and \ref{ass2}.

It remains to show that both $C_{2n}$ and $C_{3n}$ are asymptotically
negligible. Note that 
\begin{align*}
C_{2n}=& \sqrt{n}(\hat{\theta}_{n}-\theta _{0})^{\prime }\frac{1}{n}%
\sum_{i=1}^{n}\varepsilon _{i}(\theta _{0})\frac{\partial g(X_{i},\tilde{%
\theta}_{n})}{\partial \theta } \\
=& \sqrt{n}(\hat{\theta}_{n}-\theta _{0})^{\prime }\mathbb{E}\left[
\varepsilon (\theta _{0})\frac{\partial g(X,\theta _{0})}{\partial \theta }%
\right] +o_{p}(1) \\
=& o_{p}(1),
\end{align*}%
where the first equality follows by MVT, the second equality by ULLN of \cite%
{Newey1994c}, and the last step by Assumptions \ref{ass1} and \ref{ass2} as
well as the law of iterated expectations under $H_{0}$.

On the other hand, for the term $C_{3n}$, we get 
\begin{align*}
\sqrt{n}C_{3n}=& -\sqrt{n}(\hat{\theta}_{n}-\theta _{0})^{\prime }\frac{1}{n}%
\sum_{i=1}^{n}\frac{\partial q(X_{i},\tilde{\theta}_{n})}{\partial \theta }%
\frac{\partial g(X_{i},\tilde{\theta}_{n})}{\partial \theta ^{\prime }}\sqrt{%
n}(\hat{\theta}_{n}-\theta _{0}) \\
=& -\sqrt{n}(\hat{\theta}_{n}-\theta _{0})^{\prime }\mathbb{E}\left[
g(X,\theta _{0})\frac{\partial g(X,\theta _{0})}{\partial \theta ^{\prime }}%
\right] \sqrt{n}(\hat{\theta}_{n}-\theta _{0})+o_{p}(1) \\
=& O_{p}(1),
\end{align*}%
following similar arguments in proving the negligibility of $C_{2n}$. Hence $%
C_{3n}=O_{p}(n^{-1/2})=o_{p}(1)$. This ends the proof of Lemma \ref{lemmaA4}%
. $\square $

The next two lemmas establish the (uniform) convergence of $G_{n}(u,\hat{%
\theta}_{n})$ and $\Delta _{n}^{-1}(\hat{\theta}_{n})$ to $G(u,\theta _{0})$
and $\Delta ^{-1}(\theta _{0})$, respectively.

\begin{lemma}
\label{lemmaA5} Under Assumptions \ref{ass1}-\ref{ass3}, we have 
\begin{equation*}
\sup_{u\in \Pi }\left\vert G_{n}(u,\hat{\theta}_{n})-G(u,\theta
_{0})\right\vert =o_{p}(1).
\end{equation*}
\end{lemma}

\noindent \textbf{Proof of Lemma \ref{lemmaA5}:} The proof follows directly
from the ULLN of \cite{Newey1994c}. $\square $

\begin{lemma}
\label{LemmaA6} Under Assumptions \ref{ass1}-\ref{ass2}, we have 
\begin{equation*}
\Delta _{n}^{-1}(\hat{\theta}_{n})=\Delta ^{-1}(\theta _{0})+o_{p}(1).
\end{equation*}
\end{lemma}

\noindent \textbf{Proof of Lemma \ref{LemmaA6}:} The proof follows from the
ULLN of \cite{Newey1994c} and the continuous mapping theorem. $\square
\bigskip $

Now, we are ready to proceed with the proofs of our main theorems.

\noindent \textbf{Proof of Theorem \ref{thh0}:} By a straightforward
decomposition, we have 
\begin{align*}
\hat{R}_{n}^{p}(u)=& \frac{1}{\sqrt{n}}\sum_{i=1}^{n}\varepsilon _{i}(\hat{%
\theta}_{n})\left( 1\left\{ q(X_{i},\hat{\theta}_{n})\leq u\right\}
-g^{\prime }(X_{i},\hat{\theta}_{n})\Delta _{n}^{-1}(\hat{\theta}%
_{n})G_{n}(u,\hat{\theta}_{n})\right) \\
=& \hat{R}_{n}(u)-G_{n}^{\prime }(u,\hat{\theta}_{n})\Delta _{n}^{-1}(\hat{%
\theta}_{n})\frac{1}{\sqrt{n}}\sum_{i=1}^{n}\varepsilon _{i}(\hat{\theta}%
_{n})g(X_{i},\hat{\theta}_{n}) \\
:=& \hat{R}_{n}(u)-G_{n}^{\prime }(u,\hat{\theta}_{n})\Delta _{n}^{-1}(\hat{%
\theta}_{n})\hat{S}_{n}.
\end{align*}%
By Lemmas \ref{lemmaA3}-\ref{LemmaA6}, we have that 
\begin{align*}
\hat{R}_{n}^{p}(u)=& \frac{1}{\sqrt{n}}\sum_{i=1}^{n}\varepsilon _{i}(\theta
_{0})1\left\{ q(X_{i},\theta _{0})\leq u\right\} -G^{\prime }(u,\theta _{0})%
\sqrt{n}(\hat{\theta}_{n}-\theta _{0}) \\
& -G^{\prime }(u,\theta _{0})\Delta ^{-1}(\theta _{0})\left[ \frac{1}{\sqrt{n%
}}\sum_{i=1}^{n}\varepsilon _{i}(\theta _{0})g(X_{i},\theta _{0})-\Delta
(\theta _{0})\sqrt{n}(\hat{\theta}_{n}-\theta _{0})\right] +o_{p}(1) \\
=& \frac{1}{\sqrt{n}}\sum_{i=1}^{n}\varepsilon _{i}(\theta _{0})\left(
1\left\{ q(X_{i},\theta _{0})\leq u\right\} -G^{\prime }(u,\theta
_{0})\Delta ^{-1}(\theta _{0})g(X_{i},\theta _{0})\right) +o_{p}(1) \\
=& R_{n0}^{p}(u)+o_{p}(1),
\end{align*}%
uniformly in $u\in \Pi $.

The weak convergence of $R_{n0}^{p}(u)$ and consequently the weak
convergence of $\hat{R}_{n}^{p}(u)$ to the centered Gaussian process $%
R_{\infty}^{p}$ with covariance structure $K^p(u_1,u_2)$ in (\ref{kw}) can
be readily obtained by showing that the finite-dimensional distributions of $%
R_{n0}^p(u)$ converge to those of $R_{\infty}^{p}$ and the asymptotic
equicontinuity of $R_{n0}^{p}(u)$ by a direct application of Lemma \ref%
{lemDonsker}. This ends the proof of Theorem \ref{thh0}. $\square$

\noindent \textbf{Proof of Corollary \ref{corho}:} The weak convergence of
the empirical process $\hat{R}_{n}^{p}(u)$ and the continuous mapping
theorem ensure the convergence in distribution of $\Gamma (\hat{R}_{n}^{p})$
to $\Gamma (R_{\infty }^{p})$ for any continuous functional $\Gamma (\cdot )$
and in particular that of $KS_{n}$ to $KS_{\infty }$.

For the test statistic $CvM_{n}$, we will prove that 
\begin{equation*}
\int_{\Pi }\left\vert \hat{R}_{n}^{p}(u)\right\vert ^{2}\,F_{n}(du)%
\xrightarrow{d}\int_{\Pi }\left\vert R_{\infty }^{p}(u)\right\vert
^{2}\,F_{\theta _{0}}(du).
\end{equation*}%
The weak convergence of the processes $\hat{R}_{n}^{p}(u)$ and $\sqrt{n}%
(F_{n}(u)-F_{\theta _{0}}(u))$ (by Lemma \ref{lemDonsker} and Assumption \ref{ass1}) and the Skorohod construction (see 
%TCIMACRO{\TeXButton{Serfling1980}{\citealp{Serfling1980}}}%
%BeginExpansion
\citealp{Serfling1980}%
%EndExpansion
), yield 
\begin{equation}
\sup_{u}\left\vert \hat{R}_{n}^{p}(u)-R_{\infty }^{p}(u)\right\vert
\rightarrow _{a.s.}0,  \label{sk1}
\end{equation}%
and 
\begin{equation}
\sup_{u}\left\vert F_{n}(u)-F_{\theta _{0}}(u)\right\vert \rightarrow
_{a.s.}0.  \label{sk2}
\end{equation}%
Now write 
\begin{align*}
\left\vert \int_{\Pi }\left\vert \hat{R}_{n}^{p}(u)\right\vert
^{2}\,F_{n}(du)-\int_{\Pi }\left\vert R_{\infty }^{p}(u)\right\vert
^{2}\,F_{\theta _{0}}(du)\right\vert \leq & \left\vert \int_{\Pi }\left(
\left\vert \hat{R}_{n}^{p}(u)\right\vert ^{2}-\left\vert R_{\infty
}^{p}(u)\right\vert ^{2}\right) \,F_{n}(du)\right\vert \\
& +\left\vert \int_{\Pi }\left\vert R_{\infty }^{p}(u)\right\vert
^{2}\,\left( F_{n}(du)-F_{\theta _{0}}(du)\right) \right\vert.
\end{align*}%
The first term of the right-hand side of the above inequality is $o(1)$ $a.s.$ due to (\ref{sk1}). The trajectories of the limiting process $R_{\infty
}^{p}(u) $ are bounded and continuous almost surely. Then, by applying
Helly-Bray Theorem (see p.97 in 
%TCIMACRO{\TeXButton{Rao1965}{\citealp{Rao1965}}}%
%BeginExpansion
\citealp{Rao1965}%
%EndExpansion
) to each of these trajectories and taking into account (\ref{sk2}), we
obtain $\left\vert \int_{\Pi }\left\vert R_{\infty }^{p}(u)\right\vert
^{2}\,\left( F_{n}(du)-F_{\theta _{0}}(du)\right) \right\vert \rightarrow
_{a.s.}0$. This concludes the proof of Corollary \ref{corho}. $\square $

\noindent \textbf{Proof of Theorem \ref{thh1}:} Under Assumptions \ref{ass1}-%
\ref{ass3}, uniformly in $u\in \Pi $, 
\begin{align*}
& \sup_{u\in \Pi }\left\vert \frac{1}{n}\sum_{i=1}^{n}\left\{ \varepsilon
_{i}(\hat{\theta}_{n})\mathcal{P}_{n}1\left\{ q(X_{i},\hat{\theta}_{n})\leq
u\right\} -\mathbb{E}\left[ \varepsilon (\theta _{0})\mathcal{P}1\left\{
q(X,\theta _{0})\leq u\right\} \right] \right\} \right\vert \\
=& \sup_{u\in \Pi }\left\vert \frac{1}{\sqrt{n}}\hat{R}_{n}^{p}(u)-\mathbb{E}%
\left[ \left( p\left( X\right) -q\left( X,\theta _{0}\right) \right) 
\mathcal{P}1\left\{ q(X,\theta _{0})\leq u\right\} \right] \right\vert \\
=& o_{p}(1)
\end{align*}%
by ULLN of \cite{Newey1994c} and similar arguments in proving Lemmas \ref%
{lemmaA2}, \ref{lemmaA5} and \ref{LemmaA6}. $\square $

\noindent \textbf{Proof of Theorem \ref{thh1n}:} Note that under the local
alternatives $H_{1n}$ in (\ref{h1n}), we have that uniformly in $u\in \Pi $: 
\begin{align*}
\hat{R}_{n}^{p}(u)=& \frac{1}{\sqrt{n}}\sum_{i=1}^{n}\left( \varepsilon _{i}(%
\hat{\theta}_{n})-\frac{r(q(X_{i},\hat{\theta}_{n}))}{\sqrt{n}}\right) 
\mathcal{P}_{n}1\left\{ q(X_{i},\hat{\theta}_{n})\leq u\right\} \\
& +\frac{1}{n}\sum_{i=1}^{n}r(q(X_{i},\hat{\theta}_{n}))\mathcal{P}%
_{n}1\left\{ q(X_{i},\hat{\theta}_{n})\leq u\right\} \\
=& \frac{1}{\sqrt{n}}\sum_{i=1}^{n}\left( \varepsilon _{i}(\theta _{0})-%
\frac{r(q(X_{i},\theta _{0}))}{\sqrt{n}}\right) \mathcal{P}1\left\{
q(X_{i},\theta _{0})\leq u\right\} \\
& +\mathbb{E}\left[ r(q(X,\theta _{0}))\mathcal{P}1\left\{ q(X,\theta
_{0})\leq u\right\} \right] +o_{p}(1) \\
:=& R_{n1}^{p}(u)+\Delta _{r}(u)+o_{p}(1) \\
\Rightarrow & R_{\infty }^{p}+\Delta _{r},
\end{align*}%
where the second equality follows by similar arguments in proving
Theorem 1 and by ULLN. Since $\varepsilon _{i}(\theta
_{0})-n^{-1/2}r(q(X_{i},\theta _{0}))$ forms a zero mean and $i.i.d.$ summand
in this local alternative framework, we can apply the functional central
limit theorem to $R_{n1}^{p}(u)$, just as we applied it to $R_{n0}^{p}(u)$
defined in (\ref{rn0}), leading to $R_{n1}^{p}(u)\Rightarrow R_{\infty }^{p}$%
. The last step then follows and we finish the proof of Theorem \ref{thh1n}. 
$\square $

\noindent \textbf{Proof of Theorem \ref{bootstrap}:} As in Theorem \ref{thh0}%
, we have the following decomposition: 
\begin{align*}
\hat{R}_{n}^{p\ast }(u)=& \frac{1}{\sqrt{n}}\sum_{i=1}^{n}\varepsilon _{i}(%
\hat{\theta}_{n})\left( 1\left\{ q(X_{i},\hat{\theta}_{n})\leq u\right\}
-g^{\prime }(X_{i},\hat{\theta}_{n})\Delta _{n}^{-1}(\hat{\theta}%
_{n})G_{n}(u,\hat{\theta}_{n})\right) V_{i} \\
=& \frac{1}{\sqrt{n}}\sum_{i=1}^{n}\varepsilon _{i}(\hat{\theta}%
_{n})1\left\{ q(X_{i},\hat{\theta}_{n})\leq u\right\} V_{i}-G_{n}^{\prime
}(u,\hat{\theta}_{n})\Delta _{n}^{-1}(\hat{\theta}_{n})\frac{1}{\sqrt{n}}%
\sum_{i=1}^{n}\varepsilon _{i}(\hat{\theta}_{n})g(X_{i},\hat{\theta}%
_{n})V_{i} \\
:=& \hat{R}_{n}^{\ast }(u)-G_{n}^{\prime }(u,\hat{\theta}_{n})\Delta
_{n}^{-1}(\hat{\theta}_{n})\hat{S}_{n}^{\ast }.
\end{align*}%
By Lemma \ref{lemDonsker}, it follows from a stochastic equicontinuity
argument and the consistency of $\hat{\theta}_{n}$ to $\theta_0$ that,
uniformly in $u\in \Pi $, 
\begin{equation*}
\hat{R}_{n}^{\ast }(u)=\frac{1}{\sqrt{n}}\sum_{i=1}^{n}\varepsilon
_{i}(\theta _{0})1\left\{ q(X_{i},\theta _{0})\leq u\right\} V_{i}+o_{p}(1),
\end{equation*}%
and 
\begin{equation*}
\hat{S}_{n}^{\ast }=\frac{1}{\sqrt{n}}\sum_{i=1}^{n}\varepsilon _{i}(\theta
_{0})g(X_{i},\theta _{0})V_{i}+o_{p}(1).
\end{equation*}%
Thus, by Lemmas \ref{lemmaA5} and \ref{LemmaA6}, uniformly in $u$, 
\begin{align*}
\hat{R}_{n}^{p\ast }(u)& =\frac{1}{\sqrt{n}}\sum_{i=1}^{n}\varepsilon
_{i}(\theta _{0})(1\left\{ q(X_{i},\theta _{0})\leq u\right\} -G^{\prime
}(u,\theta _{0})\Delta ^{-1}(\theta _{0})g(X_{i},\theta _{0}))V_{i}+o_{p}(1)
\\
& =\frac{1}{\sqrt{n}}\sum_{i=1}^{n}\varepsilon _{i}(\theta _{0})\mathcal{P}%
1\left\{ q(X_{i},\theta _{0})\leq u\right\} V_{i}+o_{p}(1) \\
& :=R_{n0}^{p\ast }(u)+o_{p}(1),
\end{align*}%
leading to the multiplier bootstrapped version of $R_{n0}^{p}(u)$ in (\ref%
{rn0}). The rest of the proof then follows from the multiplier central limit
theorem applied to process $R_{n0}^{p\ast }(u)$; see van der Vaart and
Wellner (1996, Theorem 2.9.2, p.179), and the continuous mapping theorem. $%
\square $

\section*{\center{Appendix B: Simulation results with trimming}}\label{appendix_b}

\renewcommand{\thesection}{B} \renewcommand{\theassumption}{B.%
\arabic{assumption}} \renewcommand{\thelemma}{B.\arabic{lemma}}

\renewcommand\thetable{B.\arabic{table}} \setcounter{table}{0} %
\setcounter{lemma}{0} \setcounter{equation}{0} \setcounter{figure}{0}

In Section \ref{MC} we found that the inverse probability weighting type $ATE$ estimators and traditional balancing tests can perform poorly when estimated
propensity scores are relatively close to zero or one. In this section, we
conduct the same Monte Carlo study as in Section \ref{MC} but we trim
observations with estimated propensity score outside $[0.05,0.95]$ when
considering $\widehat{ATE_{n}}$ in \eqref{ate} and the classical balancing tests; for the
other test statistics, we do not use any trimming. Table \ref{tab:mc1trim}
and Table \ref{tab:mc2trim} present the results under designs $DGP1$-$DGP5$,
and $DGP6$-$DGP10$, respectively.

\afterpage{
\begin{landscape}
\begin{table}[htbp]
\caption{Monte Carlo results under designs $DGP1$-$DGP5$ with trimming}
\label{tab:mc1trim}
\centering
\centering\begin{adjustbox}{ max width=1\linewidth, max totalheight=1\textheight, keepaspectratio}
\begin{threeparttable}
    \begin{tabular}{ccccccccccccccccc} \hline
    \toprule
    DGP   & $n$     & \multicolumn{1}{c}{$CvM_{n}$} & \multicolumn{1}{c}{$KS_{n}$} & \multicolumn{1}{c}{$T_{n}(0.01)$} & \multicolumn{1}{c}{$T_{n}(0.05)$} & \multicolumn{1}{c}{$T_{n}(0.10)$} & \multicolumn{1}{c}{$T_{n}(0.15)$} & \multicolumn{1}{l}{Max-$t$} & \multicolumn{1}{l}{$Wald$} & \multicolumn{1}{c}{$CvM_{n}^{unp}$} & \multicolumn{1}{c}{$KS_{n}^{unp}$} & \multicolumn{1}{c}{$CvM_{n}^{trad}$} & \multicolumn{1}{c}{$KS_{n}^{trad}$} & \multicolumn{1}{c}{Bias} & \multicolumn{1}{c}{CI length} & \multicolumn{1}{c}{Coverage} \\ \hline
    \midrule

    1     & 100   & 5.00  & 5.60  & 4.90  & 2.30  & 0.80  & 0.20  & 0.10  & 0.30  & 5.40  & 5.00  & 5.40  & 6.10  & 0.02  & 1.26  & 93.70 \\
    1     & 200   & 4.40  & 4.60  & 4.90  & 2.30  & 0.60  & 0.20  & 2.40  & 2.10  & 4.00  & 4.50  & 4.40  & 3.90  & 0.01  & 0.91  & 94.00 \\
    1     & 400   & 5.10  & 5.80  & 4.10  & 2.40  & 1.10  & 0.40  & 4.20  & 4.50  & 4.70  & 5.30  & 4.70  & 4.80  & 0.01  & 0.67  & 93.90 \\
    1     & 600   & 5.20  & 5.70  & 4.30  & 1.80  & 0.70  & 0.40  & 4.80  & 5.00  & 5.00  & 5.30  & 4.60  & 5.50  & 0.00  & 0.56  & 93.30 \\
    1     & 800   & 5.30  & 4.30  & 4.50  & 1.50  & 0.50  & 0.40  & 5.00  & 6.10  & 4.90  & 4.50  & 6.20  & 4.70  & 0.00  & 0.49  & 93.70 \\
    1     & 1000  & 5.00  & 5.90  & 4.50  & 2.90  & 1.20  & 0.80  & 6.00  & 6.80  & 5.10  & 4.90  & 5.70  & 5.00  & 0.00  & 0.44  & 93.80 \\ \hline
    2     & 100   & 28.90 & 26.40 & 6.70  & 7.60  & 7.70  & 6.20  & 14.20 & 21.40 & 17.90 & 18.70 & 16.10 & 16.80 & 0.64  & 3.30  & 76.10 \\
    2     & 200   & 61.50 & 55.30 & 7.30  & 17.30 & 20.80 & 18.70 & 12.50 & 19.70 & 41.60 & 42.60 & 35.50 & 30.50 & 0.68  & 2.62  & 75.40 \\
    2     & 400   & 90.60 & 84.30 & 18.50 & 44.60 & 50.40 & 50.40 & 13.80 & 20.00 & 79.80 & 74.80 & 71.50 & 60.60 & 0.68  & 1.94  & 70.00 \\
    2     & 600   & 98.90 & 96.30 & 33.60 & 66.60 & 74.00 & 75.80 & 19.60 & 26.10 & 94.90 & 93.40 & 89.80 & 82.80 & 0.68  & 1.63  & 59.70 \\
    2     & 800   & 99.70 & 99.10 & 50.40 & 83.50 & 89.20 & 89.70 & 22.60 & 28.60 & 99.10 & 98.00 & 97.70 & 93.80 & 0.67  & 1.43  & 51.60 \\
    2     & 1000  & 100.00 & 100.00 & 67.40 & 93.90 & 96.10 & 96.80 & 28.50 & 31.70 & 99.90 & 99.90 & 99.40 & 98.40 & 0.68  & 1.29  & 45.20 \\ \hline
    3     & 100   & 32.80 & 26.10 & 6.20  & 5.30  & 1.50  & 0.40  & 0.10  & 0.00  & 33.60 & 25.00 & 15.90 & 18.10 & 0.01  & 0.87  & 95.90 \\
    3     & 200   & 59.10 & 49.90 & 18.70 & 16.50 & 6.40  & 1.10  & 0.00  & 0.00  & 59.00 & 49.60 & 40.80 & 38.80 & -0.01 & 0.57  & 94.70 \\
    3     & 400   & 69.20 & 64.10 & 38.40 & 28.60 & 9.20  & 2.40  & 0.00  & 0.00  & 69.70 & 63.70 & 83.00 & 80.30 & 0.00  & 0.39  & 94.70 \\
    3     & 600   & 79.70 & 74.60 & 56.60 & 40.90 & 14.70 & 3.80  & 0.00  & 0.00  & 80.10 & 75.30 & 98.70 & 97.10 & 0.00  & 0.32  & 95.70 \\
    3     & 800   & 81.00 & 77.70 & 63.40 & 42.30 & 15.20 & 4.10  & 0.00  & 0.00  & 81.30 & 77.80 & 99.70 & 99.70 & 0.00  & 0.27  & 96.20 \\
    3     & 1000  & 83.30 & 81.30 & 71.00 & 48.90 & 17.70 & 3.90  & 0.10  & 0.00  & 82.90 & 81.00 & 100.00 & 100.00 & 0.00  & 0.24  & 95.80 \\ \hline
    4     & 100   & 15.20 & 12.20 & 4.40  & 1.50  & 1.10  & 0.50  & 0.10  & 0.00  & 15.60 & 12.00 & 20.00 & 11.40 & 0.16  & 0.90  & 90.80 \\
    4     & 200   & 35.00 & 27.00 & 5.10  & 7.30  & 6.50  & 3.50  & 0.50  & 0.30  & 37.50 & 26.80 & 42.00 & 25.10 & 0.16  & 0.60  & 81.20 \\
    4     & 400   & 64.70 & 53.80 & 8.10  & 17.20 & 21.10 & 17.00 & 3.60  & 1.60  & 70.30 & 51.20 & 74.60 & 46.30 & 0.16  & 0.42  & 69.30 \\
    4     & 600   & 85.30 & 75.40 & 13.40 & 33.00 & 39.00 & 35.70 & 7.80  & 4.80  & 88.30 & 72.50 & 90.50 & 69.90 & 0.16  & 0.34  & 56.10 \\
    4     & 800   & 92.50 & 86.70 & 19.00 & 50.00 & 60.40 & 59.40 & 9.40  & 7.40  & 94.80 & 85.00 & 96.70 & 83.70 & 0.16  & 0.29  & 42.00 \\
    4     & 1000  & 96.50 & 93.80 & 29.80 & 63.20 & 74.00 & 75.40 & 12.60 & 10.00 & 97.60 & 92.00 & 99.20 & 90.40 & 0.16  & 0.26  & 32.50 \\ \hline
    5     & 100   & 11.00 & 7.80  & 5.00  & 2.20  & 1.30  & 0.60  & 4.20  & 4.10  & 9.20  & 6.90  & 9.20  & 5.70  & 0.37  & 1.68  & 72.70 \\
    5     & 200   & 16.40 & 13.70 & 4.90  & 2.70  & 2.30  & 1.60  & 9.50  & 10.90 & 13.80 & 9.70  & 13.00 & 7.80  & 0.44  & 1.29  & 62.10 \\
    5     & 400   & 26.70 & 23.00 & 5.90  & 5.70  & 4.70  & 3.90  & 18.80 & 18.90 & 23.50 & 15.80 & 22.70 & 11.10 & 0.47  & 0.99  & 48.60 \\
    5     & 600   & 35.80 & 32.80 & 5.30  & 8.70  & 9.90  & 8.40  & 23.70 & 26.20 & 35.20 & 21.70 & 32.40 & 15.50 & 0.46  & 0.84  & 42.60 \\
    5     & 800   & 45.00 & 39.30 & 5.70  & 11.80 & 14.30 & 12.90 & 28.80 & 31.60 & 44.00 & 28.60 & 38.80 & 19.10 & 0.46  & 0.74  & 32.40 \\
    5     & 1000  & 56.00 & 49.20 & 7.80  & 17.00 & 20.70 & 19.30 & 35.30 & 35.70 & 57.20 & 38.30 & 52.40 & 27.00 & 0.46  & 0.66  & 23.90 \\ \hline 
        \bottomrule
    \end{tabular}    \begin{tablenotes}[para,flushleft]
\small{
Note: Simulations based on 1,000 Monte Carlo experiments. ``$CvM_{n}$'' and ``$KS_{n}$'' stand for our proposed Cram\'{e}r-von Mises and Kolmogorov-Smirnov tests. ``$T_{n}(c)$'' stands for \cite{Shaikh2009}'s test, with bandwidth $h_{n}=cn^{-1/8}$. 
``Max-$t$'' and ``$Wald$'' stand for, respectively, the Bonferroni-corrected Max-$t$-test, and Wald balancing tests based on $\widehat{ATE}_{n}\left( X^{j}\right)$ defined in (\ref{ate.x}) but with observations with estimated propensity score outside $[0.05,0.95]$ trimmed.  ``$CvM_{n}^{unp}$'' and ``$KS_{n}^{unp}$'' are the  Cram\'{e}r-von Mises and Kolmogorov-Smirnov tests based on the unprojected empirical process $\hat{R}_{n}(u)$ defined in (\ref{R1}), whereas ``$CvM_{n}^{trad}$'' and ``$KS_{n}^{trad}$'' are defined analogously, but based on $\hat{R}_{n}^{trad}\left( x\right)$ as defined in (\ref{rtrad}). Finally, ``Bias'', ``CI length', and ``Coverage'' stand for the average simulated bias, estimated 95\% confidence interval length, and 95\% coverage probability for the $ATE$ estimator $\widehat{ATE_{n}}$ as defined in (\ref{ate}), but with observations with estimated propensity score outside $[0.05,0.95]$ trimmed. All entries are proportions of rejections at 5\% level, in percentage points, except ``Bias'' ,``CI length'', and ``Coverage'' (measure in percentage points), which are as described above. See the main text for further details.}
\end{tablenotes}
\end{threeparttable}
\end{adjustbox}
\end{table}
\end{landscape}
\clearpage
}

\afterpage{
\begin{landscape}
\begin{table}[htbp]
\caption{Monte Carlo results under designs $DGP6$-$DGP10$ with trimming}
\label{tab:mc2trim}\centering
\centering\begin{adjustbox}{ max width=1\linewidth, max totalheight=1\textheight, keepaspectratio}
\begin{threeparttable}
    \begin{tabular}{ccccccccccccccccc} \hline
    \toprule
    DGP   & $n$     & \multicolumn{1}{c}{$CvM_{n}$} & \multicolumn{1}{c}{$KS_{n}$} & \multicolumn{1}{c}{$T_{n}(0.01)$} & \multicolumn{1}{c}{$T_{n}(0.05)$} & \multicolumn{1}{c}{$T_{n}(0.10)$} & \multicolumn{1}{c}{$T_{n}(0.15)$} & \multicolumn{1}{l}{Max-$t$} & \multicolumn{1}{l}{$Wald$} & \multicolumn{1}{c}{$CvM_{n}^{unp}$} & \multicolumn{1}{c}{$KS_{n}^{unp}$} & \multicolumn{1}{c}{$CvM_{n}^{trad}$} & \multicolumn{1}{c}{$KS_{n}^{trad}$} & \multicolumn{1}{c}{Bias} & \multicolumn{1}{c}{CI length} & \multicolumn{1}{c}{Coverage} \\ \hline
    \midrule

    6     & 100   & 8.50  & 9.40  & 5.80  & 2.50  & 1.00  & 0.20  & 0.30  & 0.30  & 7.80  & 9.40  & 1.10  & 1.70  & -0.05 & 2.94  & 95.20 \\
      6     & 200   & 5.70  & 6.10  & 4.00  & 2.50  & 1.00  & 0.40  & 0.30  & 0.10  & 5.70  & 6.50  & 2.30  & 3.90  & -0.01 & 1.80  & 95.00 \\
    6     & 400   & 5.60  & 6.00  & 5.10  & 2.40  & 0.90  & 0.30  & 0.40  & 0.00  & 5.80  & 6.30  & 3.30  & 4.50  & -0.01 & 1.22  & 95.10 \\
    6     & 600   & 5.00  & 5.20  & 4.90  & 1.70  & 0.60  & 0.20  & 1.20  & 0.10  & 3.70  & 5.00  & 3.80  & 3.70  & -0.01 & 0.98  & 93.60 \\
    6     & 800   & 5.40  & 5.20  & 5.10  & 2.70  & 1.40  & 0.70  & 1.20  & 0.40  & 4.60  & 5.00  & 2.60  & 4.20  & -0.01 & 0.84  & 94.80 \\
    6     & 1000  & 4.90  & 6.00  & 4.50  & 2.80  & 1.10  & 0.50  & 1.90  & 0.80  & 5.70  & 5.80  & 3.70  & 5.40  & 0.00  & 0.76  & 93.70 \\ \hline
    7     & 100   & 9.50  & 10.60 & 4.10  & 1.60  & 0.70  & 0.00  & 0.40  & 2.60  & 6.80  & 8.40  & 2.50  & 3.80  & 0.00  & 5.56  & 95.70 \\
    7     & 200   & 15.40 & 14.90 & 5.10  & 3.30  & 2.00  & 1.10  & 0.70  & 1.50  & 13.60 & 13.90 & 2.70  & 6.00  & 0.22  & 3.45  & 97.70 \\
    7     & 400   & 26.40 & 24.20 & 5.40  & 6.60  & 6.80  & 5.20  & 0.40  & 0.60  & 23.00 & 19.90 & 4.60  & 7.90  & 0.39  & 2.30  & 98.50 \\
    7     & 600   & 37.20 & 33.00 & 9.10  & 10.70 & 11.10 & 8.70  & 0.70  & 0.70  & 33.70 & 30.10 & 8.30  & 9.30  & 0.44  & 1.82  & 97.30 \\
    7     & 800   & 49.50 & 46.50 & 9.90  & 16.70 & 17.00 & 13.90 & 0.60  & 0.50  & 46.40 & 37.70 & 13.80 & 13.70 & 0.47  & 1.56  & 90.80 \\
    7     & 1000  & 57.20 & 52.90 & 11.90 & 22.90 & 24.30 & 21.60 & 0.50  & 0.30  & 55.60 & 47.50 & 17.70 & 16.20 & 0.50  & 1.41  & 82.70 \\ \hline
    8     & 100   & 9.50  & 10.60 & 5.70  & 1.30  & 0.80  & 0.20  & 1.30  & 5.50  & 8.20  & 9.90  & 2.20  & 3.70  & -0.12 & 7.46  & 93.00 \\
    8     & 200   & 18.30 & 18.70 & 4.40  & 4.20  & 3.60  & 2.80  & 1.40  & 5.00  & 14.80 & 15.50 & 3.40  & 5.30  & 0.21  & 4.69  & 97.30 \\
    8     & 400   & 46.50 & 41.80 & 8.10  & 16.00 & 15.80 & 11.60 & 1.10  & 3.00  & 41.70 & 37.10 & 9.10  & 9.10  & 0.49  & 3.26  & 98.80 \\
    8     & 600   & 66.30 & 62.10 & 15.80 & 27.50 & 28.10 & 24.20 & 1.70  & 2.00  & 60.60 & 54.20 & 20.80 & 16.10 & 0.59  & 2.65  & 96.30 \\
    8     & 800   & 79.90 & 74.40 & 23.30 & 42.70 & 45.60 & 40.00 & 2.10  & 2.80  & 74.90 & 68.30 & 30.90 & 22.30 & 0.67  & 2.27  & 87.60 \\
    8     & 1000  & 87.00 & 82.10 & 31.30 & 52.70 & 55.00 & 51.20 & 2.20  & 2.80  & 82.40 & 77.70 & 42.20 & 25.40 & 0.73  & 2.03  & 77.90 \\ \hline
    9     & 100   & 10.00 & 12.00 & 3.50  & 2.40  & 0.40  & 0.20  & 0.30  & 1.90  & 10.10 & 12.40 & 1.20  & 2.50  & -0.30 & 4.82  & 89.80 \\
    9     & 200   & 15.10 & 15.10 & 4.70  & 3.70  & 1.80  & 0.80  & 0.50  & 2.40  & 12.40 & 12.00 & 4.00  & 7.10  & -0.17 & 2.89  & 90.40 \\
    9     & 400   & 30.40 & 26.80 & 4.50  & 3.10  & 2.80  & 2.90  & 0.70  & 1.80  & 21.30 & 17.00 & 8.50  & 12.00 & 0.01  & 2.03  & 93.70 \\
    9     & 600   & 44.10 & 36.10 & 5.70  & 6.10  & 6.70  & 6.00  & 1.50  & 4.10  & 32.50 & 25.30 & 17.10 & 20.10 & 0.05  & 1.67  & 93.10 \\
    9     & 800   & 57.60 & 50.20 & 6.60  & 8.80  & 11.00 & 10.40 & 2.10  & 4.30  & 46.80 & 35.60 & 25.40 & 29.30 & 0.10  & 1.44  & 90.50 \\
    9     & 1000  & 70.50 & 57.50 & 5.30  & 11.30 & 15.70 & 16.60 & 1.60  & 4.70  & 59.30 & 45.60 & 35.60 & 39.80 & 0.13  & 1.31  & 89.50 \\ \hline
    10    & 100   & 7.50  & 8.30  & 4.10  & 1.30  & 0.60  & 0.30  & 0.10  & 0.10  & 6.80  & 8.70  & 4.00  & 3.90  & -0.11 & 2.24  & 97.20 \\
    10    & 200   & 10.00 & 10.10 & 5.10  & 1.80  & 0.80  & 0.30  & 0.00  & 0.00  & 10.30 & 9.30  & 4.40  & 4.90  & -0.10 & 1.25  & 96.70 \\
    10    & 400   & 19.20 & 16.70 & 5.20  & 3.80  & 2.80  & 1.80  & 0.00  & 0.00  & 20.60 & 14.90 & 6.90  & 8.60  & -0.11 & 0.81  & 93.50 \\
    10    & 600   & 36.20 & 28.10 & 4.60  & 7.30  & 6.80  & 4.80  & 0.10  & 0.00  & 37.70 & 27.70 & 10.10 & 8.20  & -0.11 & 0.65  & 91.30 \\
    10    & 800   & 50.20 & 38.90 & 7.70  & 11.30 & 13.20 & 11.10 & 0.10  & 0.00  & 52.80 & 38.80 & 15.20 & 12.60 & -0.12 & 0.55  & 87.50 \\
    10    & 1000  & 64.30 & 51.70 & 8.20  & 18.20 & 22.30 & 18.90 & 0.10  & 0.00  & 67.00 & 51.40 & 21.00 & 13.90 & -0.12 & 0.49  & 83.10 \\ \hline
        \bottomrule
    \end{tabular}    \begin{tablenotes}[para,flushleft]
\small{
Note: Simulations based on 1,000 Monte Carlo experiments. ``$CvM_{n}$'' and ``$KS_{n}$'' stand for our proposed Cram\'{e}r-von Mises and Kolmogorov-Smirnov tests. ``$T_{n}(c)$'' stands for \cite{Shaikh2009}'s test, with bandwidth $h_{n}=cn^{-1/8}$. 
``Max-$t$'' and ``$Wald$'' stand for, respectively, the Bonferroni-corrected Max-$t$-test, and Wald balancing tests based on $\widehat{ATE}_{n}\left( X^{j}\right)$ defined in (\ref{ate.x}) but with observations with estimated propensity score outside $[0.05,0.95]$ trimmed.  ``$CvM_{n}^{unp}$'' and ``$KS_{n}^{unp}$'' are the  Cram\'{e}r-von Mises and Kolmogorov-Smirnov tests based on the unprojected empirical process $\hat{R}_{n}(u)$ defined in (\ref{R1}), whereas ``$CvM_{n}^{trad}$'' and ``$KS_{n}^{trad}$'' are defined analogously, but based on $\hat{R}_{n}^{trad}\left( x\right)$ as defined in (\ref{rtrad}). Finally, ``Bias'', ``CI length', and ``Coverage'' stands for the average simulated bias, estimated 95\% confidence interval length, and 95\% coverage probability for the $ATE$ estimator $\widehat{ATE_{n}}$ as defined in (\ref{ate}), but with observations with estimated propensity score outside $[0.05,0.95]$ trimmed. All entries are proportions of rejections at 5\% level, in percentage points, except ``Bias'' ,``CI length'', and ``Coverage'' (measure in percentage points), which are as described above. See the main text for further details.}
\end{tablenotes}
\end{threeparttable}
\end{adjustbox}
\end{table}
\end{landscape}
\clearpage
}

As discussed in the main text, when the propensity score is correctly
specified ($DGP1$ and $DGP6$), the IPW estimator $\widehat{ATE_{n}}$ has
very attractive finite sample properties, with little to no bias, coverage
probability close to its nominal level, and 95\% confidence interval length
shrinking with sample size $n$. Here it is important to emphasize that in $%
DGP1$, in contrast with the ``untrimmed'' $ATE$ estimator in Table \ref%
{tab:sim1}, the coverage probability of the ``trimmed'' $ATE$ estimator
under is close to the nominal level. However, when the propensity score is
misspecified, bias does not vanish as sample size increases, and coverage
probability can be substantially lower than its nominal level.

In terms of size, we note that under $DGP1$ balancing tests seem to perform
relatively well once we trim observations with ``extreme'' propensity score
estimates, though with $n=1,000$, the Wald test seems to be over-rejecting.
On the other hand, when the dimension of the covariates is relatively high
as in $DGP6$, these balancing tests tend to be conservative even with
trimming, reflecting the ``curse of dimensionality''. In terms of power, it
is clear that balancing tests are dominated by our projection-based tests in
all designs. In particular, as shown in Table \ref{tab:mc2trim} balancing
tests can have zero to no power when the dimension of covariates is high.
The comparison between our tests and other specification tests is exactly as
discussed in the main text and is therefore omitted. 

\end{appendices}

\section*{\center {Acknowledgements}}
We would like to thank Han Hong and Elie Tamer (Editors), an Associate Editor, and
four anonymous referees for comments that greatly improved this paper. We
also thank Daniel Millimet for sharing with us the data we use in the
empirical application, Tymon S\l {}oczy\'nski, Tatsushi Oka, and several
seminar and conference participants for their useful comments.

Pedro H. C. Sant'Anna acknowledge financial support from the Spanish Plan
Nacional de I+D+I (Grant No. ECO2014-55858-P).  Xiaojun Song acknowledge financial support from the
National Natural Science Foundation of China (Grant No. 71532001) and Key Laboratory of Mathematical Economics and Quantitative Finance (Peking University), Ministry of Education.

%\pagebreak 
\onehalfspacing{\small 
\bibliographystyle{jasa}
\bibliography{pscore}
}

\end{document}